\documentclass[titlepage,pallis,twoside]{wpallis}
\usepackage{fancyh}
\usepackage[l]{floatflt}
\usepackage{epsfig}
\usepackage{amssymb}
\usepackage{latexsym}
\usepackage{times}
\usepackage{slashed,cite}
\usepackage{upgreek}
\numberwithin{equation}{section}
\usepackage{amsmath}
\usepackage{amsfonts}
\usepackage{amsbsy}
\usepackage{amscd}
\usepackage{bbm}
\usepackage{multirow,bigstrut}

\newcommand{\crefs}[1]{Refs.~\cite{#1}}

\newcommand{\bal}{\begin{align}}
\newcommand{\eal}{\end{align}}
\newcommand{\beqs}{\begin{subequations}}
\newcommand{\eeqs}{\end{subequations}}
\newcommand{\eec}{\end{center}}
\newcommand{\bec}{\begin{center}}

\newcommand{\eem}{\end{matrix}}
\newcommand{\bem}{\begin{matrix}}
\newcommand{\eeq}{\end{equation}}
\newcommand{\beq}{\begin{equation}}
\newcommand{\ba}{\begin{array}}
\newcommand{\ea}{\end{array}}
\newcommand{\bea}{\begin{eqnarray}}
\newcommand{\eea}{\end{eqnarray}}
\newcommand{\baq}{\begin{eqnarray}}
\newcommand{\eaq}{\end{eqnarray}}

\newcommand\eqs[2]{Eqs.~(\ref{#1}) and (\ref{#2})}

\newcommand\eqss[3]{Eqs.~(\ref{#1}), (\ref{#2}) and (\ref{#3})}

\newcommand{\ftn}{\footnotesize}

\newcommand{\GeV}{{\mbox{\rm GeV}}}

\newcommand{\sFref}[2]{Fig.~\ref{#1}-{\ftn\sf ({#2})}}

\newcommand{\Srefs}[1]{Secs.~\ref{#1}}

\newcommand{\etal}{{\it et al.\/}}

\def\to{\rightarrow}

\def\llgm{\left\lgroup}
\def\rrgm{\right\rgroup}
\def\lf{\left(}
\def\rg{\right)}
\newcommand\vev[1]{\langle {#1} \rangle}

\newcommand{\Vhi}{\ensuremath{V_{\rm HI}}}
\newcommand{\Hhi}{\ensuremath{H_{\rm HI}}}

\newcommand{\Whi}{\ensuremath{W}}

\newcommand{\Vhio}{\ensuremath{ V_{\rm HI0}}}
\newcommand{\Ns}{\ensuremath{{N_\star}}}
\newcommand{\tNs}{\ensuremath{{\widetilde N_\star}}}

\newcommand{\mP}{\ensuremath{m_{\rm P}}}

\newcommand{\mbl}{\ensuremath{M_{BL}}}
\newcommand{\bl}{\ensuremath{B-L}}
\newcommand{\mgut}{\ensuremath{M_{\rm GUT}}}

\newcommand{\Ggut}{\ensuremath{G_{B-L}}}
\newcommand{\Gbl}{\ensuremath{U(1)_{B-L}}}

\newcommand{\lda}{\ensuremath{\lambda}}
\newcommand{\ldb}{\ensuremath{\lambda'}}

\def\openone{\leavevmode\hbox{\small1\kern-3.8pt\normalsize1}}
\newcommand{\dV}{\ensuremath{\Delta V_{\rm HI}}}

\newcommand{\fr}{\ensuremath{f_K}}
\newcommand{\ca}{\ensuremath{c_{\Phi}}}

\newcommand{\msn}{\ensuremath{\what m_{\rm \dph}}}

\newcommand{\ks}{\ensuremath{k_\star}}
\newcommand{\ns}{\ensuremath{n_{\rm s}}}
\newcommand{\na}{\ensuremath{{N_{a}}}}
\newcommand{\nb}{\ensuremath{N_S}}

\newcommand{\as}{\ensuremath{a_{\rm s}}}
\newcommand{\As}{\ensuremath{A_{\rm s}}}
\newcommand{\rw}{\ensuremath{r_{0.002}}}
\newcommand{\rs}{\ensuremath{\delta_\lambda}}
\newcommand{\rcc}{\ensuremath{\mathcal{R}}}
\newcommand{\rce}{\ensuremath{{\mathcal{R}}}}
\newcommand{\Ve}{\ensuremath{{V}}}

\newcommand{\sni}{\ensuremath{N^c_i}}

\newcommand{\Dex}{\ensuremath{\Delta_{\rm \star}}}

\newcommand{\kb}{\ensuremath{{K_2}}}
\newcommand{\kaa}{\ensuremath{{K_{(11)^2}}}}
\newcommand{\kba}{\ensuremath{{K_{21}}}}
\newcommand{\tkaa}{\ensuremath{{\widetilde K_{(11)^2}}}}
\newcommand{\tkba}{\ensuremath{{\widetilde K_{21}}}}

\newcommand{\kbaa}{\ensuremath{{K_{2(11)^2}}}}
\newcommand{\kbba}{\ensuremath{{K_{221}}}}
\newcommand{\tkbaa}{\ensuremath{{\widetilde K_{2(11)^2}}}}
\newcommand{\tkbba}{\ensuremath{{\widetilde K_{221}}}}

\newcommand{\dphi}{\ensuremath{\what{\delta\phi}}}
\newcommand{\dph}{\ensuremath{\delta\phi}}

\newcommand{\what}{\ensuremath{\widehat}}
\newcommand{\wtilde}{\ensuremath{\widetilde}}

\newcommand{\mnfb}{\ensuremath{\mathcal{M}_{21}}}
\newcommand{\mnfa}{\ensuremath{\mathcal{M}_{(11)^2}}}
\newcommand{\mnfs}{\ensuremath{\mathcal{M}_{S}}}
\newcommand{\ml}{\ensuremath{M_{\ld}}}

\def\al{{\alpha}}

\def\n{\bar{n}}

\def\th{{\theta}}
\def\vth{{\vartheta}}

\def\thb{{\bar\theta}}

\def\thn{{\theta_{\Phi}}}

\newcommand{\bJ}{\ensuremath{\bar J}}

\newcommand{\phc}{\ensuremath{\Phi}}
\newcommand{\phcb}{\ensuremath{\bar\Phi}}

\newcommand{\Trh}{\ensuremath{T_{\rm rh}}}
\newcommand{\sg}{\ensuremath{\phi}}

\newcommand{\sgx}{\ensuremath{\phi_\star}}
\newcommand{\sgf}{\ensuremath{\phi_{\rm f}}}

\newcommand{\ld}{\ensuremath{\lambda}}
\newcommand{\ldu}{\ensuremath{\uplambda}}

\newcommand{\kp}{\ensuremath{\kappa}}

\newcommand{\se}{\ensuremath{\widehat \phi}}
\newcommand{\sex}{\ensuremath{\widehat{\phi}_\star}}
\newcommand{\sef}{\ensuremath{\widehat{\phi}_{\rm f}}}
\newcommand{\geu}{\ensuremath{ g}}
\newcommand{\eph}{\ensuremath{\epsilon}}
\newcommand{\ith}{\ensuremath{\eta}}

\newcommand\mtta[4]{\mbox{
$\llgm\bem #1 &#2 \cr #3& #4\eem\rrgm$}}

\newcommand\mttb[9]{\mbox{
$\llgm\bem #1 &#2&#3 \cr #4&#5&#6  \cr #7&#8&#9\eem\rrgm$}}

\def\trns{transplanckian}
\def\Ka{K\"{a}hler potential}
\def\Km{K\"{a}hler manifold}
\def\Kaa{K\"{a}hler~}
\def\sub{subplanckian}

\def\bcp{{\sc\small Bicep2}/{\it Keck Array}}
\newcommand{\plk}{{\it Planck}}
\newcommand{\ma}{{MI}}
\newcommand{\mb}{{MII}}
\newcommand{\tmd}{{TM$_4$}}

\newcommand{\diag}{\ensuremath{{\sf diag}}}

\newcommand{\cU}{\ensuremath{\mathcal{U}}}
\newcommand{\ch}{\ensuremath{\mathcal{P}}}

\begin{document}
\thispagestyle{empty}


%

\title[]{\Large\boldmath\bfseries\scshape Pole-Induced Higgs Inflation
With Hyperbolic K\"ahler Geometries}

\author{\large\bfseries\scshape  C. Pallis}
\address[] {\sl Laboratory of Physics, Faculty of
Engineering, \\ Aristotle University of Thessaloniki, \\ GR-541 24
Thessaloniki, GREECE \\ {\ftn\sl  e-mail address: }{\ftn\tt
kpallis@gen.auth.gr}}


\begin{abstract}{{\bfseries\scshape Abstract} \\
\par We present novel realizations of Higgs inflation within Supergravity
which are largely tied to the existence of a pole of order two in
the kinetic term of the inflaton field. This pole arises due to
the selected \Ka s which parameterize the $(SU(1,1)/U(1))^2$ or
$SU(2,1)/(SU(2)\times U(1))$ manifolds with scalar curvatures
${\cal R}_{(11)^2}=-4/N$ or ${\cal R}_{21}=-3/N$ respectively. The
associated superpotential includes, in addition to the Higgs
superfields, a stabilizer superfield, respects the gauge and an
$R$ symmetries and contains the first allowed nonrenormalizable
term. If the coefficient of this term is almost equal to that of
the others within about $10^{-5}$ and $N=1$, the inflationary
observables can be done compatible with the present data and the
scale $M$ of gauge-symmetry breaking may assume its value within
MSSM. Increasing $M$ beyond this value, though, inflation may be
attained with less tuning. Modifications to the \Ka s associated
with the manifolds above allow for inflation, realized with just
renormalizable superpotential terms, which results to higher
tensor-to-scalar ratios as $N$ approaches its maximum at
$N\simeq40$.}
\\ \\
{\ftn \sf Keywords: Cosmology of Theories Beyond the Standard
Model, Supergravity Models};\\
{\ftn \sf PACS codes:  98.80.Cq, 11.30.Qc, 12.60.Jv, 04.65.+e}
\\[0.2cm]
\publishedin{{\sl  J. Cosmol. Astropart. Phys.} {\bf 05}, 043
(2021)}
\end{abstract} \maketitle

\setcounter{page}{1} \pagestyle{fancyplain}


\rhead[\fancyplain{}{ \bf \thepage}]{\fancyplain{}{\sl
Pole-Induced HI with Hyperbolic K\"ahler Geometries}}
\lhead[\fancyplain{}{\sl \leftmark}]{\fancyplain{}{\bf \thepage}}
\cfoot{}

\tableofcontents{}\vspace*{-1.3cm}\noindent\rule\textwidth{.4pt}

\section{Introduction}\label{intro}

The plethora of recent \cite{plcp, plin, gws, gwsnew} and
forthcoming \cite{bcp3, prism,bird} data on the \emph{cosmic
microwave background (radiation)} ({\sf \ftn CMB}) are stimulating
a new wave of inflationary model-building \cite{review, nsreview}.
One of the central task for the successful embedding of an
inflationary model within Particle Physics is the identification
of the inflaton -- i.e., the scalar field causing inflation --
with one of the fields already present in the fundamental theory.
According to an economical and intriguing set-up, the inflaton
could play, at the end of its inflationary evolution, the role of
a Higgs field \cite{old, kaloper, sm, sm1, gut1, nmh,jhep,nmhk,
univ, uvh, ighi} leading to a spontaneous breaking of a gauge
group. The relevant models may be called \emph{Higgs inflation}
({\sf \ftn HI}) for short. Let us clarify here that with this term
we do not restrict our attention only to the inflationary models
\cite{sm, sm1} which exclusively utilize the electroweak Higgs
field as higgsflaton \cite{kaloper}.

Working in the context of \emph{Supergravity} ({\ftn\sf SUGRA}) we
find it technically convenient to exemplify HI taking as reference
theory an ``elementary" \emph{supersymmetric} ({\sf\ftn SUSY})
\emph{Grand Unified Theory} ({\sf \ftn GUT}) based on the gauge
group $G_{B-L}= G_{\rm SM}\times U(1)_{B-L}$ -- where ${G_{\rm
SM}}= SU(3)_{\rm C}\times SU(2)_{\rm L}\times U(1)_{Y}$ is the
gauge group of the \emph{Standard Model} ({\ftn\sf SM}), $B$ and
$L$ denote baryon and lepton number respectively. Actually, this
is the minimal extension of \emph{Minimal SUSY SM} ({\sf\ftn
MSSM}) which is obtained by promoting to local the already
existing $U(1)_{B-L}$ global symmetry of the SM. The spontaneous
breaking of $U(1)_{B-L}$ requires the introduction of a conjugate
pair of Higgs fields $\phc$ and $\phcb$ which may contain the
inflaton. Despite its simplicity, $G_{B-L}$ is strongly motivated
by the neutrino physics and leptogenesis. Indeed, within $\Ggut$
three right-handed neutrinos, $\sni$, are necessary to cancel the
$B - L$ gauge anomaly. Subsequently, the breaking of $\Gbl$ via
the \emph{vacuum expectation values} ({\sf\ftn v.e.vs}) of $\phc$
and $\phcb$ naturally provides large Majorana masses to the
$\sni$'s, which lead to the tiny neutrino masses via the seesaw
mechanism. Furthermore, the out-of-equilibrium decay of $\sni$
gives rise to a robust baryogenesis scenario via non-thermal
leptogenesis -- see, e.g., \crefs{nmh,ighi,uvh,univ}.

The Higgs mechanism can be implemented by higgsflaton at a scale
$M$, well below the the reduced Planck scale $\mP$, adopting the
following superpotential \cite{fhi, jean}
\beq W=\lda S\lf \bar\Phi\Phi-M^2/2\rg/2-\ldb
S(\bar\Phi\Phi)^2,\label{Whi} \eeq
where $\lda, \ldb$ and $M$ are free parameters. Besides $G_{B-L}$,
$W$ respects a continuous $R$ symmetry \cite{fhi} under which $S$
and $\Whi$ are equally charged whereas the factor $\phcb\phc$ is
uncharged.  This type of $W$ is well-known from the models of
standard \cite{fhi} (for $\ldb=0$) or shifted (for $\ldb\neq0$)
F-term hybrid inflation which utilizes $S$ as inflaton with the
$\phcb-\phc$ system being stabilized at zero, for $\ldb=0$
\cite{fhi}, or even non-zero, for $\ldb\neq0$ \cite{jean}, v.e.v
during inflation. In the case of HI, though, the roles of fields
are interchanged. The inflaton is included in the $\phcb-\phc$
system whereas $S$ is placed at the origin stabilizing the
inflationary potential to be positive and breaking SUSY via its
F-term. For these reasons it is called \cite{rube} ``stabilizer"
or Goldstino superfield. Its stability w.r.t the inflationary
perturbations can be assured if we include the following term
\cite{su11}
\beq K_{2}=N_S\ln\left(1+|S|^2/N_S\right)\label{K2}\eeq
with $0<N_S<6$ in the total \Ka. $K_2$ parameterizes the compact
manifold $\mnfs= SU(2)/U(1)$. Alternative stabilization methods of
$S$ are proposed in \cref{lee,rube} whereas such a necessity is
evaded if we assume that $S$ is a nilpotent superfield \cite{nil}
-- for inflation with a single chiral superfield see
\cref{single}.

In a series of recent papers \cite{jhep, uvh, ighi, nmhk,univ}, we
demonstrate how we can realize HI within SUGRA taking $\ldb=0$ in
\Eref{Whi} and assuming the existence a (positive) non-minimal
coupling of the higgsflaton to gravity. E.g., In \cref{nmh,uvh} we
adopt exclusively a strong form of this coupling, in
\cref{jhep,nmhk} we include also a stronger non-minimal kinetic
mixing in the inflationary sector and in \cref{ighi} we also apply
the hypothesis of induced gravity \cite{igi, r2, ig}. All the HI
models above are fully compatible with the present data
\cite{plin}, most of them do not suffer from any problem with
perturbative unitarity \cite{uvh,ighi,univ}, despite the presence
of a strong coupling in the $K$'s, and some of those
\cite{nmhk,univ,jhep} yield observable gravitational waves.
Although quite compelling, though, they are relied to \Kaa frames
which do not enjoy an exact symmetry and so, their predictability
is somehow reduced.

In this paper, we attempt to cure the handicap above, abandoning
the hypothesis of the non-minimal gravitational coupling of the
higgsflaton. In our present proposal, the inflationary period is
predominantly attained due to the existence of a pole of order two
in the kinetic term of the inflaton, prior to switching to the
canonical variables \cite{tmodel, eno5, eno7, eno19}. This pole
can be traced out in the hyperbolic geometry \cite{alinde, tkref,
sky, 7disk} of the internal space. The importance of this kind of
pole in achieving inflationary solutions is stressed in \cref{prl,
ketov} and generalized in \cref{terada, pole, pole1}. The
resulting models can be characterized as T \cite{tmodel,alinde} or
E \cite{class, phenoAt} models depending on the shape of their
potential expressed in terms of the canonically normalized
inflaton. In particular, the potential of T models features two
symmetric plateaus away from the origin whereas that of E models
develops just one shoulder. Independently of the details of the
potential, both models share common predictions regarding the
scalar spectral index $\ns$ and the tensor-to-scalar ratio $r$ as
functions of the number $\Ns$ of $e$-foldings elapsed after the
pivot scale left the inflationary horizon. Indeed, in the small
$\alpha$ limit we obtain \cite{plin}
\beq \ns=1-2/\Ns\>\>\>\mbox{and}\>\>\>
r=12\alpha/N_{\star}^2\,.\label{nstm}\eeq

Trying to realize, within this regime, HI based on the F-terms --
for D-term pole HI see \cref{dtermp} --  we select one of the
following \Ka s for the Higgs superfields
\beq
K_{(11)^2}=-N\ln\left(1-2|\phc|^2)(1-2|\phcb|^2\right)\>\>\>\mbox{or}\>\>\>
K_{21}=-N\ln\left(1-|\phc|^2-|\phcb|^2\right)^2,\label{Ks}\eeq
which respect the symmetries of $\Whi$ and parameterize
correspondingly the non-compact hyperbolic spaces -- cf.
\crefs{zwirner, eno7, eno19, class} \beq \label{mnfs} \mnfa=\lf
SU(1,1)/U(1)\rg^2\>\>\>\mbox{or}\>\>\>\mnfb=SU(2,1)/(SU(2) \times
U(1))\,, \eeq with moduli-space scalar curvatures respectively --
see Appendix~A \beq \label{Rs} {\cal
R}_{(11)^2}=-{4}/{N}\>\>\>\mbox{and}\>\>\>{\cal
R}_{21}=-{3}/{N}\,.\eeq The emergent model can be characterized as
an extension of the T model based on the quartic potential with
$N=3\alpha/2$, symbolized henceforth for short as \tmd. Although
extensively analyzed, \tmd\ is mostly realized by a gauge singlet
superfield with \Ka\ parameterizing \cite{sky,tkref, alinde}, in
Poincar\'e disc coordinates, the $SU(1,1)/U(1)$ \Km\ and not
$\mnfa$. On the other hand, $\mnfb$ is widely applied in
constructing models of Starobinsky-like inflation \cite{eno5,
eno7, eno19, eno7n, class, phenoAt, nsreview} and it has not been
previously employed in building \tmd.

Since the $K$'s in \Eref{Ks} control, besides the inflaton kinetic
term, the SUGRA inflationary potential, $\Vhi$, too the pole
appearing in the kinetic term is generically expected to appear
also in $\Vhi$. The successful implementation of HI, however, is
facilitated eliminating the pole from $\Vhi$. This aim can be
accomplished by two alternative strategies -- cf. \crefs{alinde,
tkref}:

\begin{itemize}

\item Constraining $N$ in \Eref{Ks} and tuning the coefficient of
non-renormalizable term in \Eref{Whi} such that the denominator
including the pole in $\Vhi$ is (almost) cancelled out. Indeed,
fixing $N=1$ and constraining the ratio $\ldb/\lda$ or,
equivalently, $\rs$ defined as
\beq \ldb/\lda\simeq 1+\rs\,, \label{rs}\eeq
we obtain inflationary solutions which cover the observationally
favored $\ns$ region maintaining $r$ at a low level. Variation of
$M$ as a function of $\rs$ does not disrupt the realization of HI.

\item Complicating the \Ka s so that they influence just the
kinetic part of the inflationary sector and not at all $\Vhi$. In
particular, we may adopt the following $K$'s
\beq \wtilde
K_{(11)^2}=-N\ln\frac{(1-2|\phc|^2)(1-2|\phcb|^2)}{(1-2\phcb\phc)(1-2\phcb^*\phc^*)}\>\>\>\mbox{or}\>\>\>\wtilde
K_{21}=-N\ln\frac{(1-|\phc|^2-|\phcb|^2)^2}{(1-2\phcb\phc)(1-2\phcb^*\phc^*)},\label{tKs}\eeq
which yield the same lagrangian kinetic terms and share the same
curvatures with $\kaa$ and $\kba$ respectively. It is obvious,
though, that along the inflationary path defined by the condition
$|\phc|=|\phcb|$, the $\wtilde K$'s above, considered as exponents
of the exponential factor entering the SUGRA scalar potential --
see \Sref{sugra} below --, reduce to identity and so, the
expression of $\Vhi$ coincides with its SUSY version. In these
cases, the nonrenormalizable part of $W$ in \Eref{Whi} is totally
irrelevant whereas $N$ may be handled as a free parameter yielding
thereby, in its large regime, $r$ close to its upper bound
\cite{plin}. Note, in passing, that $K$ similar to $\tkaa$ is
already employed in \cref{koitci} within a scenario where $\Gbl$
is replaced by a global $U(1)$ Peccei-Quinn symmetry.

\end{itemize}

Summarizing, in this paper we analyze two novel models of HI, \ma\
and \mb, employing $W$ in \Eref{Whi} in conjunction with $K_2$ in
\Eref{K2} and the ``untilded" or the ``tilded" $K$'s in
\eqs{Ks}{tKs} respectively. More specifically, \ma\ and \mb\ are
defined as follows
\beq\bem \label{ms} \mbox{MI:}&
\hspace*{-0.1in}K=\kbaa=\kb+\kaa\>\>\mbox{or}\>\>
K=\kbba=\kb+\kba&\hspace*{-0.05in}\mbox{with}\>\>N=1&\hspace*{-0.05in}\mbox{and}\>\>\>W\>\>\mbox{with}\>\>\ldb\neq0;\\
\mbox{MII:}&\hspace*{-0.1in}K=\tkbaa=\kb+\tkaa\>\>\mbox{or}\>\>
K=\tkbba=\kb+\tkba&\hspace*{-0.08in}\mbox{with~~free
$N$}&\hspace*{-0.06in}\mbox{and}\>\>W\>\>\mbox{with}\>\>\ldb=0.\eem\eeq
In both models $M$ in \Eref{Whi} remains a free parameter which
may be constrained by the gauge coupling unification within MSSM.

The rest of the paper is organized as follows: In Sec.~\ref{sugra}
we shortly review the derivation of the potential employed for
$B-L$ HI within SUGRA. In Sec.~\ref{fhi} our inflationary models
are confronted with observations. Our conclusions are summarized
in Sec.~\ref{con}. Some mathematical notions related to the
geometric structure of the \Km s encountered in our set-up are
exhibited in Appendix~A. Throughout, the complex scalar components
of the various superfields are denoted by the same superfield
symbol, charge conjugation is denoted by a star ($^*$) -- e.g.,
$|Z|^2=ZZ^*$ -- the symbol $,Z$ as subscript denotes derivation
\emph{with respect to} ({\ftn\sf w.r.t}) $Z$, and we use units
where the reduced Planck scale $\mP = 2.43\cdot 10^{18}~\GeV$ is
equal to unity.



\section{Supergravity Framework}\label{sugra}

In Sec.~\ref{sugra1} we describe the embedding of $B-L$ HI within
SUGRA, and then, in \Sref{sugra2}, we determine the canonically
normalized fields which are involved in our scenario. In
\Sref{sugra3}, we derive the inflationary potential of our models
in the tree level and check its stability in \Sref{sugra4}. For
presentation purposes, the formulae of this section is derived for
an unspecified $N$ value, even for MI where we set $N=1$ at last.

\subsection{General Framework}\label{sugra1}

The part of the Einstein-frame action within SUGRA related to the
complex scalars $Z^I=S,\phcb,\phc$ -- which are involved in the
$W$ and $K$'s of our models as defined in \Eref{ms} -- has the
form \cite{gref}
\beq\label{action}  {\sf S}=\int d^4x \sqrt{-{
\mathfrak{g}}}\lf-\frac{1}{2}\rcc +G_{I\bJ} \geu^{\mu\nu}D_\mu Z^I
D_\nu Z^{*\bJ}-\Ve\rg\,, \eeq
where $\rce$ is the space-time Ricci scalar curvature,
$\mathfrak{g}$ is the determinant of the background
Friedmann-Robertson-Walker metric, $g^{\mu\nu}$ with signature
$(+,-,-,-)$ and $G$ with \beqs\beq
G(Z^I,Z^{*\bJ})=K(Z^I,Z^{*\bJ})+\ln\left|W(Z^I)\right|^2\label{G}\eeq
is the \Kaa-invariant function, subject to the notation
\beq G_{I\bJ}=G_{,Z^I Z^{*\bJ}}=K_{I\bJ}\>\>\>\mbox{and}
\>\>\>G^{I\bJ}G_{L\bJ}=\delta^I_L.\label{gab}\eeq\eeqs
Also, $D_\mu$ is the gauge covariant derivative which operates on
the fields $Z^I$ as follows
\beq D_\mu S=\partial_\mu S,\>\>D_\mu \phc=\partial_\mu
\phc+igA_{BL\mu} \phc\>\>\>\mbox{and}\>\>\>D_\mu
\phcb=\partial_\mu \phcb-igA_{BL\mu} \phcb,\label{Dmu}\eeq
where $A_{BL\mu}$ is the $\Gbl$ gauge field with an associated
gauge coupling constant $g$. Here we take into account that $S$ is
a $\Gbl$ singlet whereas the $\phc$ and $\phcb$ are the $\Gbl$
Higgs fields with $(B-L)$ charges $1$ and $-1$ respectively.

Finally, $\Ve$ in \Eref{action} is the SUGRA potential which
includes the contributions $V_{\rm F}$ and $V_{\rm D}$ from F and
D terms respectively. It can be calculated via the formula
\beq \Ve=V_{\rm F}+V_{\rm D}\>\>\>\mbox{with}\>\>\> V_{\rm
F}=e^{K}\left(K^{I\bJ}D_I W D^*_{\bJ} W^*-3{\vert
W\vert^2}\right)\>\>\>\mbox{and}\>\>\> V_{\rm D}=\frac{g^2}2 {\rm
D}_{BL}^2,\label{Vsugra} \eeq
where $D_I$ is the K\"{a}hler-covariant derivative and ${\rm
D}_{BL}$ is the D term corresponding to $\Gbl$. Also, we use the
shorthand notation \beqs\beq \label{Kinv} K^{I\bJ}K_{I\bar
L}=\delta^{\bJ}_{\bar L},\>\>D_I W=W_{,Z^I}
+K_{,Z^I}W\>\>\>\mbox{and}\>\>\>{\rm D}_{BL}=Z_I\lf B-L\rg
K^I\>\>\>\mbox{with}\>\>\> K^{I}={K_{,Z^I}}.\eeq In $V_{\rm D}$ we
neglect any contribution from Fayet-Iliopoulos terms and we adopt
a trivial gauge kinetic function. For the $K$'s in \eqs{Ks}{tKs}
${\rm D}_{BL}$ takes the form
\beq \label{Dbl} {\rm D}_{BL}=
2N\lf|\phc|^2-|\phcb|^2\rg\cdot\begin{cases}
%
(1-2|\phcb|^2)^{-1}(1-2|\phcb|^2)^{-1}&\mbox{for}\>\>\>K=\kaa, \tkaa,\\
\lf1-|\phc|^2-|\phcb|^2\rg^{-1}&\mbox{for}\>\>\>K=\kba,
\tkba\,.\end{cases}\eeq\eeqs From this result we deduce that
$V_{\rm D}$ can be eliminated during HI if $|\phc|=|\phcb|$, i.e.,
if we identify inflaton with the radial parts of $\phc$ and
$\phcb$. More explicitly, we parameterize the fields of our models
as follows \cite{nmh,nmhk,univ,uvh,jhep,ighi}
\beq\label{hpar} \Phi={\sg} e^{i\th}\cos\thn,\>\>\>\bar\Phi={\sg}
e^{i\thb}\sin\thn\>\>\>\mbox{with}\>\>\>0\leq\thn\leq{\pi}/{2}\>\>\>\mbox{and}\>\>\>S=
(s +i\bar s)/{\sqrt{2}}\eeq
and investigate the implementation of HI driven by the real field
$\sg$ along the field configuration
\beq\label{inftr} \bar
s=s=\thb=\th=0\>\>\>\mbox{and}\>\>\>\thn={\pi/4}\,.\eeq
This selection ensures the D flatness, since ${\rm D}_{BL}=0$ --
cf. \cref{dtermp} --, and the avoidance of a possible runaway
problem, since the term $-3{\vert W\vert^2}$ in \Eref{Vsugra}
vanishes thanks to the constraint $S=0$ \cite{rube}.

\subsection{Canonically Normalized Fileds}\label{sugra2}

To specify the canonically normalized fields, we note that, for
all choices of $K$ in Eqs.~(\ref{Ks}) and (\ref{tKs}), $K_{I\bJ}$
along the configuration in \Eref{inftr} takes the form
\beq \label{Kab} \lf K_{I\bJ}\rg=\diag\lf
M_{\phcb\phc},K_{SS^*}\rg,\eeq where $K_{SS^*}=K_{2SS^*}=1$ and
$M_{\phcb\phc}=\lf K_{I\bar J}\rg$ is the matrix containing the
elements of $K_{I\bar J}$ with $I=\phcb, \phc$. It is found to be
\beq M_{\phcb\phc}=\begin{cases}
\kp\,\diag\lf1,1\rg&\mbox{for}\>\>\>K=\kaa, \tkaa, \\
\lf\kp\sg^2/2\rg\mtta{2/\sg^2-1}{1}{1}{2/\sg^2-1}\>\>\>&\mbox{for}\>\>\>K=\kba,
\tkba, \end{cases}\>\>
\mbox{with}\>\>\>\begin{cases}\kp=2N/\fr^{2}\\\mbox{and}\\
\fr=1-\sg^2\,.\end{cases} \label{Mk}\eeq We observe that
$M_{\phcb\phc}$ is diagonal for $K=\kaa$ and $\tkaa$ whereas it
requires diagonalization for $K=\kba$ and $\tkba$. In the latter
case its eigenvalues are
\beq \label{kpm}\kp_+=\kp\>\>\>\mbox{and}\>\>\>
\kp_-=\kp\fr\,.\eeq The equality of the kinetic terms for
``tilded" and ``untilded" $K$'s with the same arithmetic indices
can be understood by the observation that the structure of the
``tilded" $K$'s is
\beq \label{tKK} \wtilde K_{(ij)}(\phcb,\phcb^*,\phc,\phc^*) =
K_{(ij)}(\phcb,\phcb^*,\phc,\phc^*)+ K_{\rm H}(\phcb,\phc)+K_{\rm
A}(\phcb^*,\phc^*),\eeq
where $(ij)=(11)^2$ or $(ij)=21$ and the subscripts ``H" and ``A"
stand for ``holomorphic" and ``antiholomorphic" respectively. More
explicitly, comparing \Eref{tKK} with \eqs{Ks}{tKs} we infer
\beq  K_{\rm H}=N\ln(1-2\phcb\phc)\>\>\>\mbox{and}\>\>\> K_{\rm
A}=N\ln(1-2\phcb^*\phc^*).\eeq
Therefore, the mixed derivatives of $\wtilde K_{(ij)}$ coincide
with the corresponding ones of $K_{(ij)}$ since $\partial_{\bar
I}K_{\rm H}=\partial_{I}K_{\rm A}=0$ for $I=\phcb, \phc$. As a
further consequence, the \Kaa metrics $K_{I\bJ}$ and $\wtilde
K_{I\bJ}$ for $K=K_{(ij)}$ and $\wtilde K=\wtilde K_{(ij)}$ with
fixed $(ij)$ turn out to be equal.

Expanding the second term of the \emph{right-hand side} ({\sf\ftn
r.h.s}) of \Eref{action} along the path in \Eref{inftr} for
$I=\phc,\phcb$ and substituting there \Eref{Mk}, we obtain
\beqs\beq \label{kzz} K_{I\bJ}\dot Z^I \dot
Z^{*\bJ}=\begin{cases}{\kp}\dot \sg^2+\kp\sg^2\lf\dot\theta^2_+
+\dot\theta^2_-+2\dot\theta^2_\Phi
\rg/2\>\>\>&\mbox{for}\>\>\>K=\kaa, \tkaa,
\\ {\kp_+}\lf\dot \sg^2+\sg^2\dot\theta^2_+/2
\rg+{\kp_-\sg^2}\lf\dot\theta^2_-/2 +\dot\theta^2_\Phi
\rg\>\>\>&\mbox{for}\>\>\>K=\kba, \tkba,
\end{cases}\eeq
where $\th_{\pm}=\lf\bar\th\pm\th\rg/\sqrt{2}$ and the dot denotes
derivation w.r.t the cosmic time, $t$. Comparing the expressions
above with the following one
\beq K_{I\bJ}\dot Z^I \dot Z^{*\bJ} \simeq \frac12\lf\dot{\widehat
\sg}^2+\dot{\widehat \th}_+^2+\dot{\widehat \th}_-^2+\dot{\widehat
\th}_\Phi^2\rg, \label{kzzn}\eeq
we can specify the canonically normalized fields, denoted by hat,
in terms of the initial (unhatted) ones -- recall that $S$ is
already cationically normalized. Thanks to \eqs{Mk}{kpm} we can
establish for all $K$'s a unique expression regarding the
normalized inflaton , i.e.,
\beq \label{Je} {d\se}/{d\sg}=J=2\sqrt{N}/\fr,\eeq
from which we infer that the kinetic term of $\sg$ exhibits a pole
at $\sg=1$ of order two -- see $\fr$ in \Eref{Mk}. For the
remaining fields of the $\phcb-\phc$ system we find
\beq \label{Je1} \>\>\>\>\>\>\>\>\>\>\>\bem
 \begin{array}{ll}\widehat{\theta}_\pm
={\sqrt{\kp}}\sg\theta_\pm,\>\>\widehat \theta_\Phi =
\sqrt{2\kp}\sg\lf\theta_\Phi-{\pi}/{4}\rg
\>\>\>&\mbox{for}\>\>\>K=\kaa, \tkaa\,,\\ \widehat{\theta}_+
=\sqrt{\kp_+}\sg\theta_+,\>\>\widehat{\theta}_-
=\sqrt{{\kp_-}}\sg\theta_-,\>\>\widehat \theta_\Phi =
\sqrt{2\kp_-}\sg\lf\theta_\Phi-{\pi}/{4}\rg
\>\>\>&\mbox{for}\>\>\>K=\kba, \tkba\,.\end{array}\eem\eeq\eeqs
Integrating \Eref{Je} we can identify $\se$ in terms of $\sg$, as
follows
\beq
\se=\sqrt{N}\ln\frac{1+\sg}{1-\sg}\>\>\>\Rightarrow\>\>\>\sg=\tanh\frac{\se}{2\sqrt{N}},\label{se}\eeq
where we set the relevant constant of integration equal to zero.
We can easily conclude that for $0\leq\sg\leq0.999$, $\se$ gets
increased from $0$ to $7.6\sqrt{N}$, i.e., $\se$ can be much
larger than $\sg$ facilitating thereby the attainment of HI (which
is actually of chaotic type and necessitates $\se\gg1$) with \sub\
$\sg$'s. The last fact is imperative for a meaningful approach to
SUGRA. Indeed, terms of the form $(\phc\phcb)^p$ with $p>2$ in
$\Whi$ are not disallowed by the symmetries and so stabilization
of our scenario against corrections from those $W$ terms dictates
\sub\ values for $\bar\Phi$ and $\Phi$ or, via \Eref{hpar}, $\sg$.

\subsection{Inflationary Potential}\label{sugra3}

Along the trough in \Eref{inftr} the only surviving term in
\Eref{Vsugra} is
\beqs\beq \label{Vhi0}\Vhi =e^{K}K^{SS^*}\, |W_{,S}|^2. \eeq
Taking into account \eqs{K2}{ms}, we obtain
$K^{SS^*}=1/K_{2SS^*}=(1+|S|^2/\nb)^2\left.\right|_{S=0}=1$. Also,
\eqs{Ks}{tKs} yield
\beq \label{eK} e^{K}=\begin{cases}
\fr^{-2N}&\mbox{for}\>\>\>K=\kaa\>\>\>\mbox{and}\>\>\>\kba,\\
1& \mbox{for}\>\>\>K=\tkaa\>\>\>\mbox{and}\>\>\>\tkba,
\end{cases}\eeq\eeqs
i.e., the pole in $\fr$ is presumably present in $\Vhi$ of \ma,
but it disappears in $\Vhi$ of \mb, as anticipated below
\Eref{tKs}. Substituting the results above and \Eref{Whi} into
\Eref{Vhi0}, this takes its master form
\beq\Vhi=\frac{\lda^2}{16}\cdot\begin{cases}
%
{\lf\sg^2-(1+\rs)\sg^4-M^2\rg^2}/{\fr^{2N}}&\mbox{for \ma},\\
\lf\sg^2-M^2\rg^2&\mbox{for \mb}\,.\end{cases}\label{Vhi}\eeq
From the first equation above we easily infer that the elimination
of the pole from the denominator of $\Vhi$ can be implemented as
$M$ and $\rs$ tend to zero only for $N=1$. No $N$ dependence
arises for \mb.

To directly compare our models with the \tmd\ \cite{alinde, plin},
we express $\Vhi$ in \Eref{Vhi} as a function of $\se$, taking
advantage of \Eref{se}. We find
\beq \label{Vhie} \Vhi= \frac{\ld^2}{16}\cdot\begin{cases}
%
\cosh^{4N}(\se/2\sqrt{N}){\lf\tanh^2(\se/2\sqrt{N})-(1+\rs)\tanh^4(\se/2\sqrt{N})-M^2\rg^2}&\mbox{for \ma},\\
\lf\tanh^2(\se/2\sqrt{N})-M^2\rg^2&\mbox{for
\mb}\,.\end{cases}\eeq
From the expressions above we conclude that \ma\ for $N=1$ and
$M=\rs=0$ and \mb\ for $M=0$ share the same expression for $\Vhi$,
which is actually the potential adopted for \tmd\
\cite{tmodel,alinde}, i.e. \beq \label{Vhio} \Vhio=
{\lda^2}\sg^4/{16}={\lda^2}\tanh^4(\se/2)/{16}\,.\eeq  The naive
expectation that this model is observationally ruled out by now
\cite{plin} since $\Vhio=\Vhio(\sg)$ coincides with the one of the
quartic power-low model is not correct, since $\sg$ is not
canonically normalized. This is related to $\se$ via \Eref{se} and
so, the emerging picture is radically different. Indeed,
$\Vhio=\Vhio(\se)$  in \Eref{Vhio} is encountered in \tmd\ which,
although constrained, is still observationally \cite{plin} alive.


\begin{figure}[!t]\vspace*{-.12in}
\hspace*{-.19in}
\begin{minipage}{8in}
\epsfig{file=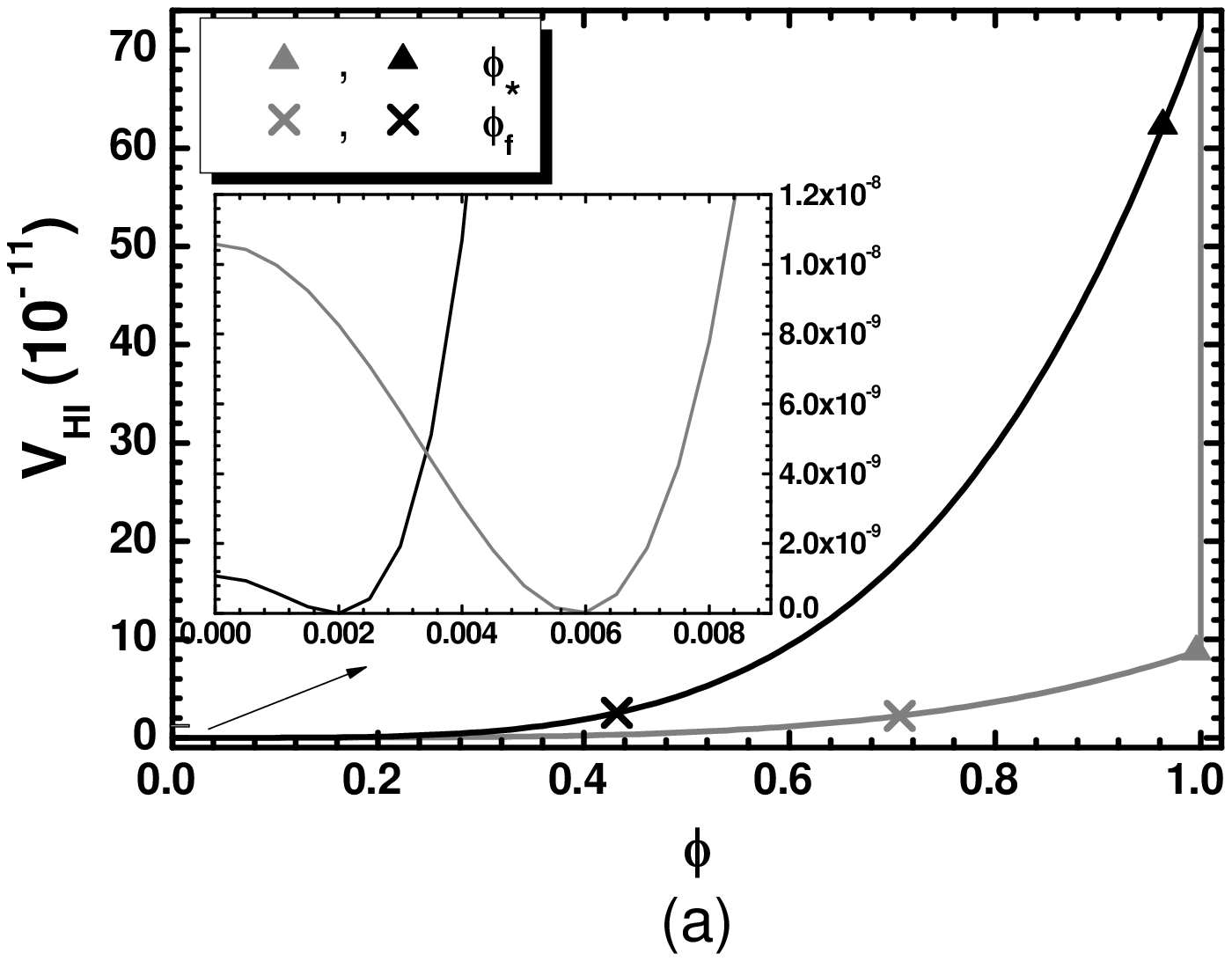,height=3.6in,angle=-90}
\hspace*{-1.2cm}
\epsfig{file=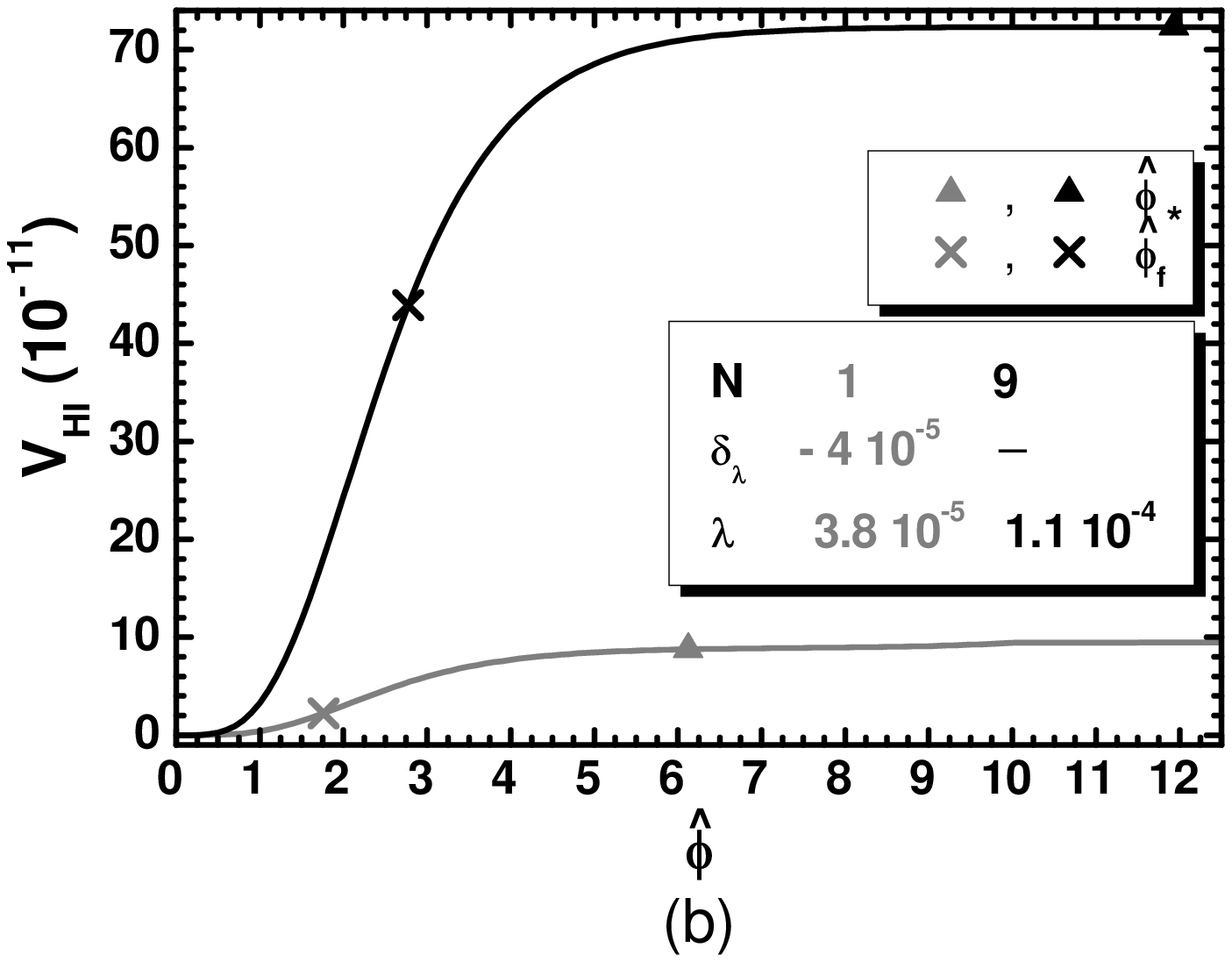,height=3.6in,angle=-90} \hfill
\end{minipage}
\hfill \vchcaption[]{\sl\small Inflationary potential $\Vhi$ as a
function of {\sffamily\ftn (a)} $\sg$ for $\sg>0$ and
{\sffamily\ftn (b)} $\se$ for $\se>0$ fixing $\mbl=\mgut$. We
consider MI with $\rs=-4\cdot10^{-5}$ and $\ld=3.8\cdot 10^{-5}$
(gray lines) or MII with $N=9$ and $\ld=1.1\cdot10^{-4}$ (black
lines). Values corresponding to $\sgx$ and $\sgf$ {\sffamily\ftn
(a)} or $\se$ and $\sef$ {\sffamily\ftn (b)} are depicted. Shown
is also the low-$\sg$ behavior of $\Vhi$ in the inset
{\sffamily\ftn (a)}. }\label{fig2}
\end{figure}

To clarify further this key feature of our models, we
comparatively plot $\Vhi$ as a function of $\sg$ in
\sFref{fig2}{a} and $\se$ in \sFref{fig2}{b} for \ma\ (gray lines)
and \mb\ (black lines) -- cf.~\cref{tmodel}. Here we impose the
constraint of MSSM gauge coupling unification which entails
$M=5.87\cdot10^{-3}$ for MI and $M=1.96\cdot10^{-3}$ for MII --
see \Sref{fhi2} and \ref{resn}. Observational constraints
determine $\ld$ and $\rs$ for MI or $\ld$ and $N$ for MII. Their
values are shown in the legend of \sFref{fig2}{b}. We notice in
\sFref{fig2}{a} that $\Vhi$ for both models have a parabolic-like
slope for $\sg<1$ whereas \ma\ exhibits a sharp increase as $\sg$
tends to $1$ due to the pole -- see \Eref{Vhi}. On the contrary,
in \sFref{fig2}{b} $\Vhi$ experiences a stretching for $\se>1$
which results to a plateau facilitating, thereby, the
establishment of HI for $\se\gg1$.  We remark that a gap of about
one order of magnitude emerges between $\Vhi=\Vhi(\se)$ for MI and
MII which has as a consequence the elevation -- see \Sref{resn} --
of $r$ as explained in \cref{rRiotto}. The two crucial values of
$\sg$ depicted in \sFref{fig2}{a}, $\sgf$ and $\sgx$, which limit
the observationally relevant inflationary period -- see
\Sref{fhi1} -- are $0.71$ and $0.99564$ for MI or $0.43$ and
$0.96322$ for MII. These values are enhanced according to
\Eref{se}, i.e., $\sef=1.763$ and $\sex=6.12$ for MI $\sef=2.77$
and $\sex=11.93$ for MII -- see \sFref{fig2}{b}. The structure of
$\Vhi$ for low $\sg$ values, responsible for the implementation of
the Higgs mechanism, is also illustrated in the inset of
\sFref{fig2}{a}. Similar behavior is expected for
$\Vhi=\Vhi(\se)$.

\subsection{Stability and
Loop-Corrections}\label{sugra4}

To consolidate our inflationary models, we have to verify that the
inflationary direction in \Eref{inftr} is stable w.r.t the
fluctuations of the non-inflaton fields. To this end, we construct
the mass-squared spectrum of the various scalars defined in
\Eref{Je1}. Taking $\rs\simeq0$ and $M\ll\sg$ we find the
expressions of the masses squared $\what m^2_{\chi^\al}$ (with
$\chi^\al=\theta_+,\theta_\Phi$ and $S$)  and arrange them in
\Tref{tab1}. These masses are compared with the Hubble parameter
during HI, $\Hhi^2=\Vhi/3$. For $\sg\simeq\sgx\simeq1$ the exposed
formulas fairly approach the quite lengthy, exact expressions
taken into account in our numerical computation. From our findings
in \Tref{tab1} we easily confirm that $\what
m^2_{\chi^\al}\gg\Hhi^2$ for $\chi^\al=\theta_+$ and $\theta_\Phi$
-- note that $g\gg\ld$. As regards the crucial stabilization of
$S$, the negative contributions of the $N$-dependent terms are
maximized for $N=1$ and both \ma\ and \mb, and so we deduce that
the minimal value of $\what m^2_{s}(\sgx)$ is $\what
m^2_{s}(\sgx)\left.\right|_{\rm min}\simeq6\Hhi^2/\nb$ which stays
positive and heavy enough for $0<N_S<6$. As $\sg$ decreases
towards the SUSY vacuum, $\what m^2_{s}$ increases, thanks to the
term with $\sg$ in the denominators. For \mb, as $N$ increases
$m^2_{s}$ increases too since the negative contributions with $N$
in the denominators decrease. In \Tref{tab1} we also display the
masses, $M_{BL}$, of the gauge boson $A_{BL}$ -- which signals the
fact that $\Gbl$ is broken (already) during HI -- and the masses
of the corresponding fermions. Note that the unspecified eigestate
$\what \psi_\pm$ is defined as \beq \what
\psi_\pm=(\what{\psi}_{\Phi+}\pm
\what{\psi}_{S})/\sqrt{2}\>\>\>\mbox{with}\>\>\>\psi_{\Phi\pm}=(\psi_\Phi\pm\psi_{\bar\Phi})/\sqrt{2}\,,\eeq
with the spinors $\psi_S$ and $\psi_{\Phi\pm}$ being associated
with the superfields $S$ and $\bar\Phi-\Phi$. The broken generator
of $\bl$ leads to one would-be-Goldstone boson $\th_-$, ``eaten''
by the gauge boson $A_{BL}$ which becomes massive. If we take into
account $\sg$, not included in \Tref{tab1}, we can easily verify
that the degrees of freedom of the particle spectrum before and
after the spontaneous breaking of $\bl$ are equal. Similarly, the
fermionic and bosonic degrees of freedom are also equal -- cf.
\cref{jhep}. As a byproduct of the fact that $\Gbl$ is already
broken during HI, no cosmic string are produced at the end of HI
-- contrary to what happens within standard F-term hybrid
inflation \cite{fhi}.

\renewcommand{\arraystretch}{1.4}

\begin{table}
\bec {\ftn \begin{tabular}{|c||c|c||l|l|l|l|}\hline {\sc
Fields}&{\sc Eigen-}& \multicolumn{5}{|c|}{\sc Masses
Squared}\\\cline{3-7} &{\sc states}&
&\multicolumn{1}{|c|}{$K=\kbaa$}&
\multicolumn{1}{|c|}{$K=\kbba$}&\multicolumn{1}{|c|}{$K=\tkbaa$}&\multicolumn{1}{|c|}{$K=\tkbba$}\\
\hline\hline
%
2 real&$\widehat\theta_{+}$&$m_{\widehat\theta+}^2$&
\multicolumn{4}{|c|}{$3\Hhi^2$}\\\cline{4-7}
scalars&$\widehat \theta_\Phi$ &$\widehat m_{ \theta_\Phi}^2$&
{$M^2_{BL}+6\Hhi^2(1+2/N$}&
{$M^2_{BL}+6\Hhi^2(1$}&\multicolumn{2}{|c|}{As for}\\\cline{6-7}
&&&
{$-1/N\sg^2-\sg^2/N)$}&{$+1/N-1/N\sg^2)$}&$K=\kbaa$&$K=\kbba$\\\cline{4-7}
1 complex&$s, {\bar{s}}$ &$ \widehat m_{
s}^2$&\multicolumn{2}{|c|}{$6\Hhi^2(1/N_S-4(1-\sg^2)/N+N\sg^2$}&\multicolumn{2}{|c|}{$6\Hhi^2(1/N_S-2/N$}\\
scalar&&&\multicolumn{2}{|c|}{$+2(1-2\sg^2)+4\sg^2/N)$}&\multicolumn{2}{|c|}{$+1/N\sg^2+\sg^2/N)$}\\\hline
1
gauge&\multirow{2}{0.31in}{$A_{BL}$}&\multirow{2}{0.31in}{$M_{BL}^2$}&
\multicolumn{4}{|c|}{\multirow{2}{0.6in}{$4Ng^2\sg^2/\fr^2$}}\\
boson&&&\multicolumn{4}{|c|}{}\\\hline
$4$ Weyl  & $\what \psi_\pm$ & $\what m^2_{ \psi\pm}$&
\multicolumn{4}{|c|}{${3\fr^2\Hhi^2/N^2\sg^2}$}
\\\cline{4-7}
spinors&$\ldu_{BL}, \widehat\psi_{\Phi-}$&$M_{BL}^2$&\multicolumn{4}{|c|}{$4Ng^2\sg^2/\fr^2$}\\
\hline
\end{tabular}}
\end{center}
\vchcaption[]{\sl\ftn The mass-squared spectra along the
inflationary trajectory of \Eref{inftr} for various $K$'s and
$\sg\lesssim1$. To avoid very lengthy formulas we set $\rs\simeq0$
and neglect terms proportional to $M\ll\sg$.}\label{tab1}
\end{table}
\renewcommand{\arraystretch}{1.}

The derived mass spectrum can be employed in order to find the
one-loop radiative corrections, $\dV$, to $\Vhi$. Considering
SUGRA as an effective theory with cutoff scale equal to $\mP$, the
well-known Coleman-Weinberg formula -- cf.
\cref{univ,uvh,nmh,jhep,r2} -- can be employed taking into account
only the masses which lie well below $\mP$, i.e., all the masses
arranged in \Tref{tab1} besides $M_{BL}$ and $\what m_{\th_\Phi}$.
The resulting $\dV$ leaves intact our inflationary outputs,
provided that the renormalization-group mass scale $Q$, is
determined by requiring $\dV(\sgx)=0$ or $\dV(\sgf)=0$, where
$\sgx$ and $\sgf$ are the observationally relevant values of $\sg$
-- see \Sref{fhi1}. The first of the conditions above -- the
latter leads to similar $Q$ values \cite{jhep} -- yields
$Q\simeq(1.1-1.4)\cdot10^{-5}$ for MI or
$Q\simeq(1.2-6.3)\cdot10^{-5}$ for MII and the values of
parameters given in \eqs{res1}{res2} below. This determination of
$Q$ renders our results practically independent of $Q$ since these
can be derived exclusively by using $\Vhi$ in \Eref{Vhi} with the
various quantities evaluated at $Q$. Note that their
renormalization-group running is expected to be negligible because
$Q$ is close to the inflationary scale $V_{\rm
HI\star}^{1/4}\simeq (2.8-3.2)\cdot10^{-3}$ for MI or $V_{\rm
HI\star}^{1/4}\simeq (2.98-6.8)\cdot10^{-3}$ for MII corresponding
to Hubble parameters $H_{\rm HI\star}\simeq (4.5-5.9)\cdot10^{-6}$
for MI or $H_{\rm HI\star}\simeq (0.5-2.5)\cdot10^{-5}$ for MII --
all the quantities are computed for $\sg=\sgx$.

\section{Inflation Analysis}\label{fhi}

In \Srefs{resa} and \ref{resn} below we ascertain analytically and
numerically respectively, if the $\Vhi$ in \Eref{Vhi} endowed with
the $\sg$ normalization in \Eref{Je} may be consistent with a
number of constraints introduced in Sec.~\ref{const}.

\subsection{General Framework} \label{const}

The constraints imposed on our inflationary setting may be grouped
in two categories (observational and theoretical) which are
described respectively in \Srefs{fhi1} and \ref{fhi2} below.

\subsubsection{Observational Constraints}\label{fhi1}

\paragraph{(a) Number of e-foldings and Normalization of the Power Spectrum.}
The number of e-foldings $\Ns$ that the scale $\ks=0.05/{\rm Mpc}$
experiences during HI and the amplitude $\As$ of the power
spectrum of the curvature perturbations generated by $\sg$ can be
computed using the standard formulae  \cite{review}
\begin{equation}
\label{Nhi}  \Ns=\int_{\sef}^{\sex} d\se\frac{\Vhi}{\Ve_{\rm
HI,\se}}\>\>\>\mbox{and}\>\>\>\As^{1/2}= \frac{1}{2\sqrt{3}\, \pi}
\; \frac{\Vhi^{3/2}(\sex)}{\left|\Ve_{\rm
HI,\se}(\sex)\right|},\eeq
where $\sex$ is the value of $\se$ when $\ks$ crosses outside the
inflationary horizon, and $\sef$ is the value of $\se$ at the end
of HI, which can be found, in the slow-roll approximation, from
the condition
\beq{\ftn\sf
max}\left\{\eph(\se),\left|\ith(\se)\right|\right\}\simeq1,\>\mbox{where}\>\>
\eph=\frac12\left(\frac{\Ve_{\rm HI,\se}}{\Ve_{\rm
HI}}\right)^2\>\>\>\mbox{and}\>\>\> \ith={\Ve_{\rm
HI,\se\se}\over\Ve_{\rm HI}}\cdot \label{srcon} \eeq
The observables above are to be confronted with the \cite{plcp}
\beq \Ns\simeq61.5+\frac14\ln\frac{\Vhi(\sex)^2}{g_{\rm
rh*}^{1/3}\Vhi(\sef)}\>\>\>\mbox{and}\>\>\>\sqrt{\As}\simeq4.588\cdot10^{-5}\,,\label{prob}\eeq
where for the latter restriction we take into account the final
full-mission \plk\ measurements (TT, TE, EE+lowE+lensing) and
\emph{Baryon Acoustic Oscillations} ({\sf\ftn BAO}) dataset. Also,
we assumed that HI is followed in turn by a oscillatory phase with
mean equation-of-state parameter $w_{\rm rh}\simeq1/3$ -- which
corresponds to a quatric potential \cite{plin} --, radiation and
matter domination, $\Trh$ is temperature after HI, $g_{\rm rh*}$
is the energy-density effective number of degrees of freedom at
the reheat temperature $\Trh$ -- for the MSSM spectrum we take
$g_{\rm rh*}=228.75$. We observe that $\Ns$ turns out to be
independent of the explicit value of $\Trh$.

\paragraph{(b) Remaining Observables.} The remaining inflationary
observables (the spectral index $\ns$, its running $\as$, and the
tensor-to-scalar ratio $r$) are estimated through the relations:
\beq\label{ns} \ns=\: 1-6\eph_\star\ +\ 2\ith_\star,\>\>\>
\as=\:\frac23\left(4\eta_\star^2-(\ns-1)^2\right)-2\xi_\star\>\>\>\mbox{and}\>\>\>
r=16\eph_\star, \eeq
where $\xi={\Ve_{\rm HI,\se} \Ve_{\rm HI,\se\se\se}/\Ve_{\rm
HI}^2}$ and the variables with subscript $\star$ are evaluated at
$\sg=\sgx$. These observables must be in agreement with the
fitting of the \plk\ TT, TE, EE+lowE+lensing, {\sffamily\ftn BK14}
and BAO data \cite{plin,gws} with $\Lambda$CDM$+r$ model which
approximately requires at 95$\%$ \emph{confidence level} ({\sf\ftn
c.l.})
\begin{equation}  \label{nswmap}
\ns=0.967\pm0.0074\>\>\>\mbox{and}\>\>\>r\leq0.07\>\>\>\mbox{with}\>\>\>|\as|\ll0.01\,.
\end{equation}
Here, {\sf\ftn BK14} data is taken by the \bcp\ CMB polarization
experiments up to and including the 2014 observing season. The
bound on $r$ tightens to $r\leq0.06$ \cite{gwsnew}, if we include
\bcp\ data from the 2015 observing season and combine them with
2015 \plk\ data. For a direct comparison of our findings with the
obervational outputs in \cref{plin,gws}, we also compute
$\rw=16\eph(\se_{0.002})$ where $\se_{0.002}$ is the value of
$\se$ when the scale $k=0.002/{\rm Mpc}$, which undergoes
$N_{0.002}=\Ns+3.22$ e-foldings during HI, crosses the horizon of
HI -- see \Sref{resn}.

\subsubsection{Theoretical Considerations}\label{fhi2}

We may qualify better our models by taking into account three
additional restrictions of theoretical origin. In particular:

\paragraph{(a) Gauge-coupling Unification.} One of the strongest motivation of
our proposal is that $W$ in \Eref{Whi} leads not only to an
inflationary era but also to the breaking of a GUT scale symmetry.
In our introductory set-up the v.e.vs of $\phcb$ and $\phc$ break
$U_{B-L}$ down to $\mathbb{Z}^{B-L}_2$. Indeed, minimizing $\Vhi$
in \Eref{Vhi} we find that a SUSY vacuum arising after the end of
HI with
\beq \vev{S}=0 \>\>\>\mbox{and}\>\>\>
|\vev{\Phi}|=|\vev{\bar\Phi}|= \begin{cases}
\sqrt{1-\sqrt{1-4M^2}}\simeq M(1+M^2/2)&\mbox{for \ma,}\\
M&\mbox{for \mb.}\end{cases} \label{vevs} \eeq
Note that other critical points of $\Vhi$ for MI are not
accessible by our inflaton -- cf. \cref{jean}. Although
$\vev{\Phi}$ and $\vev{\bar\Phi}$ break spontaneously
$U(1)_{B-L}$, no cosmic strings  are produced at the SUSY vacuum,
since this symmetry is already broken during HI.  The
contributions from the soft SUSY breaking terms can be safely
neglected within contemporary SUSY, since the corresponding mass
scale is much smaller than $M$. They may shift
\cite{r2,univ,uvh,ighi}, however, slightly $\vev{S}$ from zero in
\Eref{vevs}.

As regards the value of $M$, it can be determined by requiring
that $\vev{\Phi}$ and $\vev{\bar\Phi}$ take the values dictated by
the unification of MSSM gauge coupling constants. In particular,
the unification scale
$\mgut\simeq2/2.433\times10^{-2}\simeq8.22\cdot10^{-3}$ is to be
identified with $M_{BL}$ -- see \Tref{tab1} -- at the SUSY vacuum,
\beq \label{Mg} \vev{M_{BL}}={2\sqrt{N}gM/
\vev{\fr}}=\mgut\>\>\Rightarrow\>\>M\simeq{\mgut}/{2g\sqrt{N}}
\>\>\>\mbox{for}\>\>\>\vev{\fr}\simeq1. \eeq
Here $g\simeq0.7$ is the value of the GUT gauge coupling constant.
However, unification at a scale above $\mgut$ is also possible
\cref{unif}. Also, $U(1)_{B-L}$ gauge symmetry does not disturb
the MSSM unification. We can, therefore, treat $M$ as a free
parameter keeping in mind that its most natural value is that
given in \Eref{Mg} -- see \Sref{resn}.

The determination of $M$ influences heavily the inflaton mass at
the vacuum, $\msn$ and induces an $N$ dependence in the results.
Indeed, the (canonically normalized) inflaton,
\beq\dphi=\vev{J}\dph\>\>\>\mbox{with}\>\>\> \dph=\phi-M
\>\>\>\mbox{and}\>\>\>\vev{J}=2\sqrt{N/\vev{\fr}}\label{dphi} \eeq
acquires mass, at the SUSY vacuum in \Eref{vevs}, which is given
by
\beq \label{msn} \msn=\left\langle\Ve_{\rm
HI,\se\se}\right\rangle^{1/2}= \left\langle \Ve_{\rm
HI,\sg\sg}/J^2\right\rangle^{1/2}\simeq\frac{\lda
M}{2\sqrt{2N}}\cdot
\begin{cases}
\sqrt{1-3M^2}&\mbox{for \ma,}\\
\vev{\fr}&\mbox{for \mb,}\end{cases}\eeq
where the last (approximate) equality for \ma\ is valid only for
$M\ll1$ -- see \Eref{vevs}.

\paragraph{(b) Effective Field Theory.} Destabilization of our
inflationary scheme may be caused by higher order
non-renormalizable terms in \Eref{Whi} and non-homogenous terms to
the $K$'s in \Eref{Ks} originated from non-perturbative
(instanton) corrections within superstring theory -- see
\cref{pole1}. To minimize the ramifications to our models from
such terms, we impose two additional theoretical constraints --
keeping in mind that $\Vhi(\sg_{\rm f})\leq\Vhi(\sg_\star)$:
\beq \label{subP}\Vhi(\sgx)^{1/4}\leq1
\>\>\>\mbox{and}\>\>\>\sgx\leq1.\eeq
The first from the inequalities above is easily satisfied in our
set-up as inferred from \Fref{fig2}. The second one is
automatically fulfilled. Indeed, by construction $K$'s in
\eqs{Ks}{tKs} with the parameterizations in \Eref{hpar} induce a
kinetic pole for $\sg_{\rm p}=1$ and so HI takes place for
$\sg<1$. Even if we multiply all the quadratic terms in $K$'s by a
factor $\ca$  -- to keep the symmetries in \Eref{mnfs} -- with
$\ca<1$, then although $\sg_{\rm p}$ seems to be translated to
\trns\ values, this is not the case. This is, because if we
perform the rescalings
\beq\label{resc1}
\Phi\to\Phi/\sqrt{\ca},\>\>\bar\Phi\to\bar\Phi/\sqrt{\ca}\>\>\>\mbox{and}\>\>\>
S\to S,\eeq
$\ca$ can be absorbed in $K$'s whereas $W$ of \Eref{Whi} remains
unaltered provided we do the redefinitions
\beq\label{resc2}
\ld\to\ca\ld,\>\>\ldb\to\ca^2\ldb\>\>\>\mbox{and}\>\>\> M\to
M/\sqrt{\ca}.\eeq
As a consequence, we obtain a relocation of the pole at $\sg_{\rm
p}=1$ and HI can be processed for subplanckian $\sg$'s as analyzed
in \Sref{resa} below.

\paragraph{(c) Tuning of the Initial Conditions.}  HI may be
implemented undoubtedly if $\sg$ starts its slow-roll below the
location of kinetic pole. The closer to pole is set $\sgx$ the
larger $\Ns$ is obtained. Therefore, a tuning of the initial
conditions is required which can be somehow quantified  -- cf.
\cref{ighi,univ,nmhk,uvh} -- defining the quantity
\beq \Dex=\left(\sg_{\rm p} - \sgx\right)/\sg_{\rm
p}\,.\label{dex}\eeq
The naturalness of the attainment of HI increases with $\Dex$.

\subsection{Analytic Results} \label{resa}

The investigation of the inflationary dynamics can be performed
employing the formulae of \Sref{const} and the expressions
$\Vhi=\Vhi(\sg)$ in \Eref{Vhi} taking advantage of $J$ in
\Eref{Je} and making use of the chain rule of derivation -- for
more details see \cref{jhep}. Following this strategy, we avoid to
employ the quite complicate expression of $\Vhi=\Vhi(\se)$ in
\Eref{Vhie}. We present our findings in \Sref{res1a} and
\ref{res2a} for \ma\ and \mb\ respectively.

\subsubsection{Model I (\ma)}\label{res1a}

Taking into account that no inflationary solutions are numerically
localized for \ma\ and $N\neq1$ we confine ourselves to $N=1$ here
and hereafter. The slow-roll parameters read
\beqs\bea\label{sr1e}&&\eph={2\sg^2}\lf\frac{1+(1+\rs)\sg^2(\sg^2-2)-M^2}
{\sg^2-\sg^4(1+\rs)\sg^4-M^2}\rg^2;\\
&&\label{sr1i}\ith\simeq\frac{(3-5\sg^2)\fr}{\sg^2}+\frac{(7\sg^2-9)\rs\sg^4+M^2(10\sg^4-17\sg^2+5)}{\fr\sg^4}\,.\eea\eeqs
The condition of \Eref{srcon} in the present case reads
\beq \sg_{\rm f}\simeq \mbox{\sf\small max} \lf1/{\sqrt{2}},
\sqrt{(9-\sqrt{21})/10}\rg\,,\label{sgf1}\eeq
i.e., $\sg_{\rm f}$ is determined due to the violation of the
$\eph$ criterion. Assuming $\sgf\ll\sgx$, $\Ns$ can be
approximately computed from \Eref{Nhi} as follows
\begin{equation}
\label{Nhi1}  \Ns\:=\int_{\sgf}^{\sgx}\, J^2\frac{\Ve_{\rm
HI}}{\Ve_{\rm HI,\sg}}d\sg\:\simeq
\frac{1}{2}\frac{\sgx^2}{f_{K\star}}\>\>\>\Rightarrow\>\>\>\sgx\simeq
\sqrt{\frac{2\Ns}{2\Ns+1}}\>\>\>\mbox{where}\>\>\>f_{K\star}=1-\sgx^2.
\end{equation}
Also any dependence on the parameters $\rs$ and $M$ can be safely
neglected. Here we explicitly display the dependence of $\Ns$ on
$\sgx$ to facilitate the understanding of the final result. It is
clear that, as $\sgx$ approaches unity, $\Ns$ increases
drastically assuring thereby the achievement of efficient HI.
Indeed, from the last expression we infer that $\sgx$ is slightly
lower than unity, as anticipated in \Sref{sugra4}. Plugging $\sgx$
from \Eref{Nhi1} into the rightmost equation in \Eref{Nhi} and
solving w.r.t $\ld$, we arrive at the expression
\beq \ld=8\pi\sqrt{3\As}\frac{1-4\ml
N_{\star}^2}{\Ns(1-2\ml\Ns)^2}\>\>\>\mbox{with}\>\>\>\ml=\rs+M^2.\label{lan1}\eeq
Enforcing \Eref{prob} on $\As$ we expect that $\ld$ is comparable
with its value in the quatric power-law model \cite{plin}. Upon
substitution of $\sgx$ into \Eref{ns} we obtain the the
observational predictions of MI which are
\beq  \ns\simeq
1-2\ml\lf9+4\Ns\rg-\frac{2}{\Ns}-\frac{48\Ns^2\ml^2}{(1-2\ml\Ns)^2},\>\>\>\>\>\>\>
r\simeq\frac{8(1-4\ml
N_{\star}^2)^2}{N_{\star}^2(1-2\ml\Ns)^2}\>\>\>\>\>\>\>\>\>\mbox{and}\>\>\>
\>\>\>\>\>\>\as\sim-\frac{3}{N_{\star}^2},\label{ns1}\eeq
where the last expression (for $\as$) yields just an order of
magnitude estimation -- the exact result is too lengthly to be
worthily presented. Obviously, for $\ml=0$, the well-known
predictions of the \tmd\ in \Eref{nstm} are recovered, i.e.,
$\ns\simeq0.963$, $\as\simeq6.3\cdot10^{-4}$ and $r\simeq0.003$
for $\Ns=55$. As we can verify numerically, the expression above
provide sufficiently accurate results within the whole allowed
region of parameters presented in \Sref{resn}.

\subsubsection{Model II (\mb)}\label{res2a}

Working along the lines of the previous section, we estimate the
slow-roll parameters for \mb\ as follows
\beq\label{sr2}\eph=\frac{2\sg^2}{N}\lf\frac{\fr}{\sg^2-M^2}\rg^2\>\>\>\mbox{and}\>\>\>
\ith=\frac{\fr}{N}\frac{5\sg^4-3(1+M^2)\sg^2+M^2}{(\sg^2-M^2)^2},
\eeq
where we keep now the $N$ dependence in the formulas. HI is over
when \Eref{srcon} is saturated at the maximal $\sg$ value, $\sgf$,
from the following two values
\beq \sg_{1\rm
f}\simeq\sqrt{4+N-\sqrt{N(N+8)}}/2\>\>\>\mbox{and}\>\>\> \sg_{2\rm
f}\simeq\sqrt{N+8-\sqrt{N^2+16N+4}}/\sqrt{10},\label{sgf2}\eeq
where $\sg_{1\rm f}$ and  $\sg_{2\rm f}$ are such that
$\eph\lf\sg_{1\rm f}\rg\simeq1$ and $\ith\lf\sg_{2\rm
f}\rg\simeq1$. For $N>3$, the latter condition overshadows the
former. Note that for $N=1$ the expressions above coincide with
those in \Eref{sgf1} since the slow-roll parameters for MI and MII
tend to the same limit for $M=\rs=0$ and $N=1$.

Assuming $\sgx\gg\sgf$ and neglecting subdominant logarithmic
contributions, we can estimate $\Ns$ from \Eref{Nhi} with result
similar to that in \Eref{Nhi1}, i.e.,
\begin{equation}
\label{Nhi2}  \Ns\simeq
\frac{N}{2}\frac{\sgx^2}{f_{K\star}}\vev{\fr}\>\>\>\Rightarrow\>\>\>\sgx\simeq
\sqrt{\frac{\Ns}{\tNs}}\>\>\>
\mbox{with}\>\>\>\begin{cases}\vev{\fr}=1-M^2,\\
\tNs=\Ns+\vev{\fr}N/2.\end{cases}
\end{equation}
Given that $\vev{\fr}\simeq1$ and $N\ll\Ns$, we expect that $\sgx$
is again a little less than the pole value. Plugging $\sgx$ from
\Eref{Nhi2} into the rightmost equation in \Eref{Nhi} and solving
w.r.t $\ld$, we find
\beq \ld\simeq8\sqrt{3N\As}\pi \vev{\fr}/\Ns,\label{lan2}\eeq
which results to similar numerical value with that found for MI.
Inserting, finally, $\sgx$ from \Eref{Nhi2} into \Eref{ns} and
expanding successively the resulting equations in series of $M$
and $1/\Ns$ we obtain
\beq \label{ns2} \ns\simeq
1-\frac{3\vev{\fr}}{\Ns}+\frac{\vev{\fr}}{\tNs},\>\>\> r\simeq
\frac{8N\vev{\fr}^2}{\Ns\tNs}\>\>\>\mbox{and}\>\>\>
\as\simeq-\lf\frac{3}{N_{\star}^{2}}-\frac1{\tNs}\rg\vev{\fr}^2\,.\eeq
From the expressions above, we can infer that the predictions of
pure \tmd\ in \Eref{nstm} are revealed for $N<10$ and
$\vev{\fr}\simeq1$ -- recall that our results can be directly
compared with those in \crefs{tmodel, alinde, 7disk} setting
$2N=3\alpha$. Taking into account the definition of $\tNs$ in
\Eref{Nhi2} and solving the second formula above w.r.t $N$ we can
also obtain the upper bound $N\lesssim35$ imposing the second
inequality in \Eref{nswmap} for $\Ns\simeq55$ -- cf. \cref{7disk}.
This is in good agreement with our numerical findings, where $\Ns$
is consistently derived  from \Eref{prob}. Numerically we also
verify generically the correctness of the simple expressions above
for $N>0.1$ and $M<0.3$.

\subsection{Numerical Results}\label{resn}

Let us recall that our inflationary scenaria depends on the
following parameters -- see \eqss{Whi}{Ks}{tKs}:
$$M,\>\lda\>\>\mbox{and}\>\>\rs\>\>\mbox{for MI, or}\>\>N\>\>\mbox{for MII}.$$
Note that $\nb$ in \Eref{K2} does not affect the inflationary
outputs, provided that $\what m_{s}^2>\Hhi^2$ for every allowed
$N$. This is satisfied when $0<\nb\leq6$, as explained in
\Sref{sugra4}. Recall that we exclusively use $N=1$ for \ma, since
no acceptable inflationary solutions are found for other $N$'s.
This $N$ value can be regarded as a prediction than as a
shortcoming of \ma. The confrontation of the aforementioned
parameters with observations is implemented as follows: Upon
substitution of $\Vhi$ from \Eref{Vhi} in \eqss{Nhi}{srcon}{ns} we
extract the inflationary observables as functions of these
parameters and $\sgx$. The last quantity and $\ld$ can be
determined by enforcing the fulfilment of \Eref{prob}, whereas
$\rs$ and $N$ largely affect the outputs on $\ns$ and $r$
respectively and are constrained by \Eref{nswmap}. The predictions
for $\msn$ follow from \Eref{msn}.

We start the presentation of our results by comparing the outputs
of our models against the observational data \cite{plin,gws} in
the $\ns-\rw$ plane -- see \Fref{fig1}. More precisely, the dark
[light] shaded contours in \Fref{fig1} depict the marginalized
joint $68\%$ [$95\%$] c.l. regions obtained by fitting the current
data \cite{plin,gws} with $\Lambda$CDM$+r$ -- see Fig. 28 of
\cref{plcp}. We draw dashed [solid] lines for \ma\ [\mb] and show
the variation of $-\rs/10^{-5}$ [$N$] along each line fixing
$\vev{\mbl}=\mgut$. Variation of $\vev{\mbl}$ -- see below --
leads to undistinguishable modifications of the displayed lines
although the values of $\rs$ and $N$ are to be readjusted. For
most of the indicated points we list in the Table of \Fref{fig1}
the corresponding field values ($\sgx$ and $\sgf$), the output
parameters ($\Dex$, $\ld$ and $M$) and the inflationary
observables. Note that, given $g(\mgut)\simeq0.7$, the imposed
constraint on $\vev{\mbl}$ determines uniquely $M$ for \ma\ since
$N=1$ whereas it allows for a $N$ dependence for \mb\ -- see
\Eref{Mg}. Recall that the variation of $\Vhi$ as a function of
$\sg$ and $\se$ is illustrated in \sFref{fig2}{a} and {\sf\ftn
(b)} respectively for \ma\ and $\rs=-4\cdot10^{-5}$ (light gray
lines) or \mb\ and $N=9$ (black lines).

\begin{figure}[!t]\vspace*{-.26in}
\begin{center}
\epsfig{file=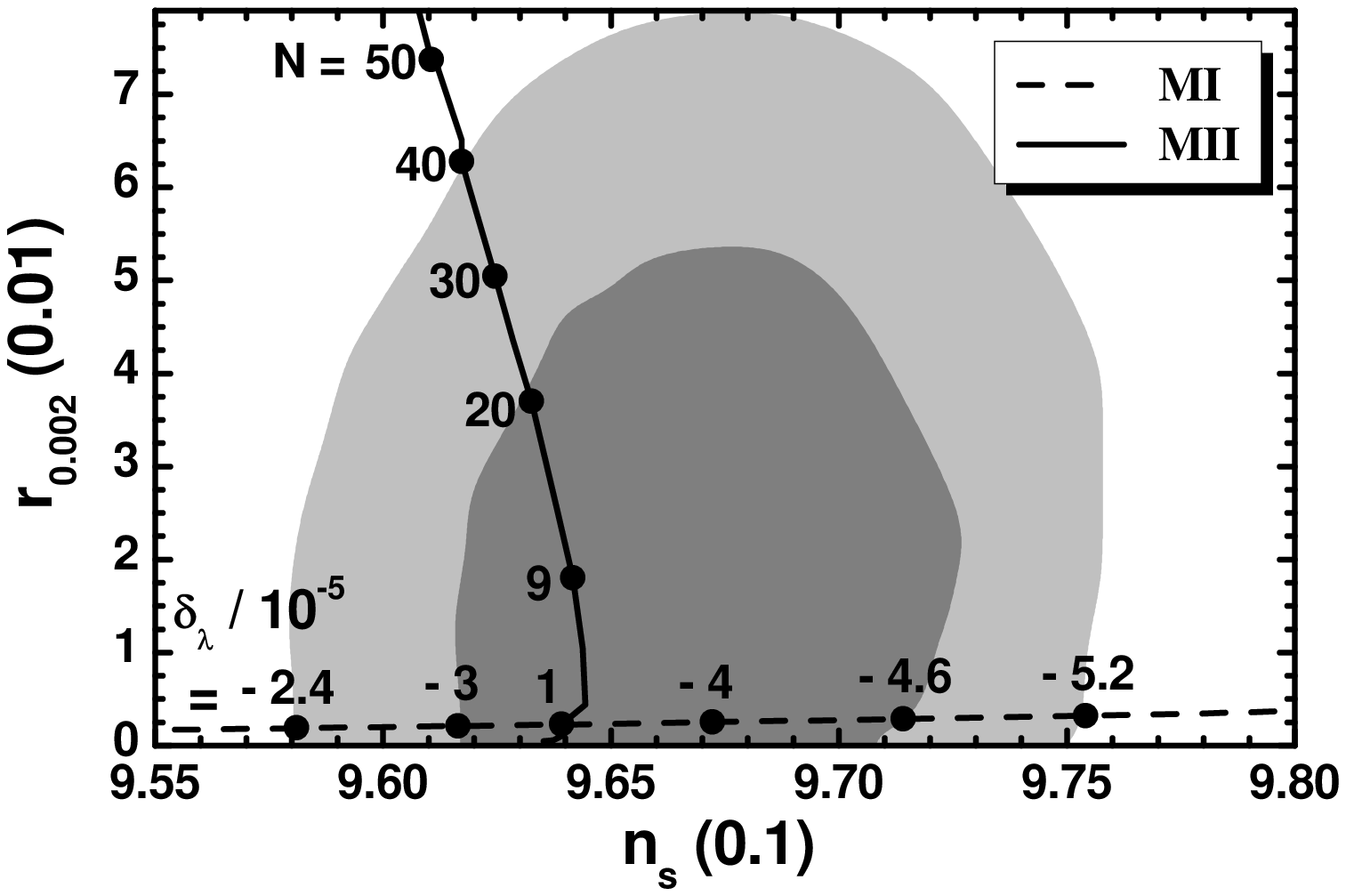,height=3.65in,angle=-90}\end{center}
\begin{center}\renewcommand{\arraystretch}{1.1}
{\ftn\begin{tabular}{|c||ccccc|ccccc|}\hline {Model:}
&\multicolumn{5}{c|}{Model I (\ma)}&\multicolumn{5}{c|}{Model II
(\mb)}\\\hline\renewcommand{\arraystretch}{1.}
${\rs/10^{-5}}$&$-2.4$&$-3$&$-4$&$-4.6$&$-5.2$&\multicolumn{5}{c|}{$-$}\\
$N$&\multicolumn{5}{c|}{$1$}&$1$&$9$&$20$&$30$&$40$\\\hline\hline
$\sgx/0.1$&$9.9538$&$9.955$&$9.9564$&$9.958$&$9.9592$&$9.9554$&$9.6322$&$9.2402$&$8.9244$&$8.64$\\
$\Dex (\%)$&$0.46$&$0.45$&$0.44$&$0.42$&$0.41$&$0.45$&$3.7$&$7.6$&$10.7$&$13.6$\\
$\sgf/0.1$&\multicolumn{5}{c|}{$7.1$}&$7.1$&$4.3$&$3.3$&$2.8$&$2.5$\\\hline
$\Ns$&$55.6$&$55.7$&$55.1$&$55.7$&$55.8$&$55.2$&$56.8$&$57.2$&$57.4$&$57.5$\\\hline
$\ld/10^{-5}$&$3.3$&$3.4$&$3.8$&$3.9$&$4.1$&$3.6$&$10.7$&$16.5$&$20.8$&$24.8$\\
$M/10^{-3}$&\multicolumn{5}{c|}{$5.87$}&$5.87$&$1.96$&$1.31$&$1.075$&$0.93$\\\hline
$\ns/0.1$&$9.58$&$9.62$&$9.67$&$9.71$&$9.75$&$9.64$&$9.64$&$9.63$&$9.62$&$9.62$\\
$-\as/10^{-4}$&$5.4$&$5.9$&$7.1$&$7.8$&$8.6$&$6.5$&$6.4$&$6.6$&$6.9$&$7$\\
$r/10^{-2}$&$0.21$&$0.23$&$0.28$&$0.31$&$0.34$&$0.25$&$2$&$4$&$5.5$&$6.9$
\\\hline
\end{tabular}}
\end{center}\vspace*{-.08in}
\hfill \vchcaption[]{\sl \small Allowed curves in the $\ns-\rw$
plane fixing $\mbl=\mgut$ for \ma\ and various $\rs$'s indicated
on the dashed line or \mb\ and various $N$'s indicated on the
solid line. The marginalized joint $68\%$ [$95\%$] c.l. regions
from \plk\ TT, TE, EE+lowE+lensing, {\sffamily\ftn BK14} and BAO
data \cite{plin} are depicted by the dark [light] shaded contours.
The relevant field values, parameters and observables
corresponding to points shown in the plot are listed in the
Table.}\label{fig1}
\end{figure}

Focusing first on \ma, we see that the whole observationally
favored range at low $r$'s is covered varying $(-\rs)$ which
remains rather close to $10^{-5}$. In other words, HI is feasible
only at the cost of a tuning of order $10^{-5}$ on $\rs$ and
another, of the order $10^{-2}$ on $\Dex$. This amount of tuning
is unavoidable, though, in this kind of models -- cf.
\cref{eno5,eno7,nsreview}. The assumption $\rs=0$ yields results
identical to \tmd\ but it causes concerns regarding the naturality
of the model since such arrangement requires the exact equality of
the coefficients of two unrelated terms in $W$. In particular, let
$\ns$ vary in the range of \Eref{nswmap} from the findings listed
in the Table of \Fref{fig1}, we obtain \beq\label{res1}
2.4\lesssim\frac{-\rs}{10^{-5}}\lesssim
5.2,\>\>\>4.6\gtrsim\frac{\Dex}{10^{-3}}\gtrsim
4.1,\>\>\>5.4\lesssim\frac{-\as}{10^{-4}}\lesssim
8.6\>\>\>\mbox{and}\>\>\> 2.1\lesssim
\frac{r}{10^{-3}}\lesssim3.4\,.\eeq Since \ma\ predicts
$r\gtrsim0.0019$, it is testable by the forthcoming experiments,
like {\sc Bicep3} \cite{bcp3}, PRISM \cite{prism} and LiteBIRD
\cite{bird}, which are expected to measure $r$ with an accuracy of
$10^{-3}$.

Turning now to \mb, we observe from \Fref{fig1} that almost all
the allowed $r$'s are accessible by varying $N$ whereas $\Dex$
increases with $N$. That is, the required tuning respecting the
initial conditions becomes less and less severe as $N$ increases.
We employ here just integer $N$'s since these values are better
motivated from the string theory. Comparing our present outputs
with those of the models in \crefs{jhep,nmhk,univ}, where similar
observable $r$ values are achieved with decimal $N$ values, we
have to recognize that the quality of the $N$ adjustment is better
here. On the other hand, $\ns$ is concentrated a little lower than
its central value in \Eref{nswmap} without possibility of further
variation without alteration to the required value of $\Ns$. From
the results accumulated in the Table of \Fref{fig1} we find the
following allowed ranges of parameters: \beq\label{res2}
0.962\lesssim\ns\lesssim0.964,\>\>\>1\lesssim N\lesssim
40,\>\>\>0.45\gtrsim{\Dex}/{10^{-2}}\gtrsim
13.6\>\>\>\mbox{and}\>\>\> 0.0025\lesssim {r}\lesssim0.07\,.\eeq
Regarding $\as$, it varies in the range $-(6.3-7.1)\cdot10^{-4}$
and so, \mb\ is also consistent with the fitting of data with the
$\Lambda$CDM+$r$ model \cite{plin}. \mb\ not only can be probed by
the aforementioned future experiments but it can also become
compatible with optimistic current central values of $r$ mentioned
in \cref{gwsnew}. E.g., we are able to achieve $r=0.02$ or
$r=0.012$ with $N=9$ or $N=5.5$ correspondingly.

As we mention in \Sref{const}, $\msn$ is affected heavily from the
choice of $M$. For the most natural $M$ choice done in
\Fref{fig1}, we obtain
\beq\label{resmsn} 1.7\lesssim {\msn/
10^{11}~\GeV}\lesssim2.1\>\>\>\mbox{or}\>\>\>1.8\lesssim
{\msn/10^{11}~\GeV}\lesssim1.9\,, \eeq
for \ma\ and \mb\ respectively. Here and hereafter we restore
units for convenience regarding the results on $\msn$. These
$\msn$ ranges let, in principle, open the possibility of
non-thermal leptogenesis if we introduce a suitable coupling
between $\bar\Phi$ and the right-handed neutrinos -- see e.g.
\crefs{ighi,uvh,univ,nmh}. This issue, though, needs further
investigation in order to be surely verified.

Taking advantage from the fact that $M$ can be liberated from any
gauge unification constraint -- see \Sref{fhi2} -- we investigate
the modification of our results if we vary $M$ continuously from a
low scale $M\simeq0.001$ up to a maximal one which corresponds to
$\vev{\mbl}=1$ with $g\simeq0.7$. The indicative low bound on $M$
stems from the fact that $\Whi$ in \Eref{Whi} is established
around $\mP$ and so we expect that the scale $M$, entered by hand
in the theory, must have comparable size. Our outputs for \ma\ and
\mb\ are presented in \sFref{fig3}{a} and {\sf\ftn (b)}
respectively and discussed below.

As regards \ma, the allowed (shaded) region in the $M-(-\rs)$
plane is bounded from the upper and lower limits on $\ns$ in
\Eref{nswmap} which are depicted by a dashed and a dot-dashed line
respectively -- see the segment of the graph enclosed in a box and
enlarged in the inset. These lines coincide with each other in the
bulk of the plot. So we draw just a solid line corresponding to
the central value of $\ns$ in \Eref{nswmap}. Along this line in
\sFref{fig3}{a} we get
\beq\label{resm1}  10^{-3}\lesssim M\lesssim0.2,\>\>\>
5.5\cdot10^{-4}\lesssim {-\rs}/10^{-2}\lesssim4\>\>\>
\mbox{and}\>\>\>3.2\lesssim {\msn}/10^{10}~\GeV\lesssim644.7 \eeq
with $\ld\sim3.7\cdot10^{-5}$, $\Dex\simeq0.0043$ and
$\as\simeq-0.00067$. For $M$ values beyond the upper bound above
we do not succeed to achieve consistent inflationary solutions. We
see that the tuning w.r.t $\rs$ is ameliorated as $M$ increases,
whereas that of $\Dex$ remains unaltered.

\begin{figure}[!t]\vspace*{-.12in}
\hspace*{-.19in}
\begin{minipage}{8in}
\epsfig{file=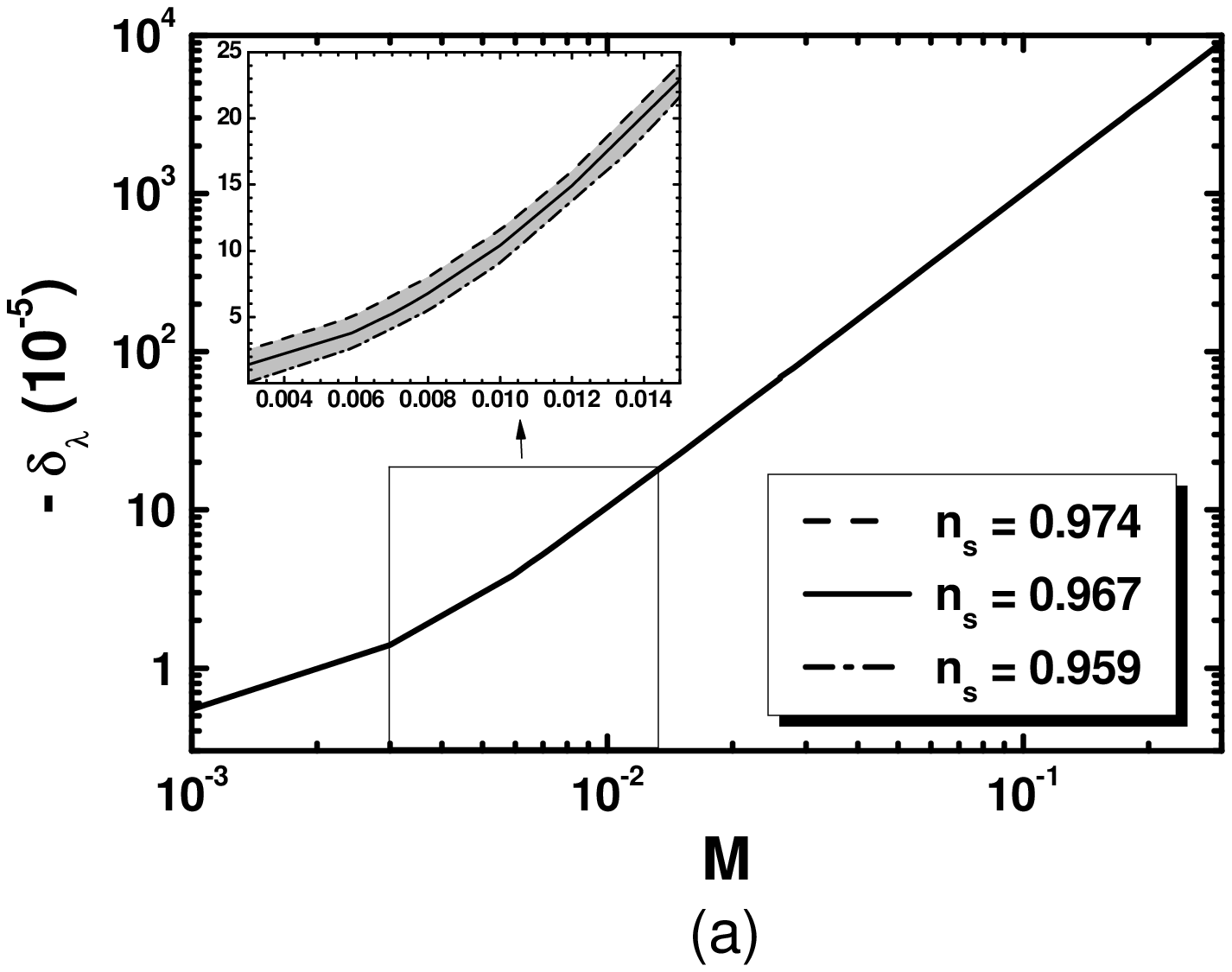,height=3.6in,angle=-90}
\hspace*{-1.2cm}
\epsfig{file=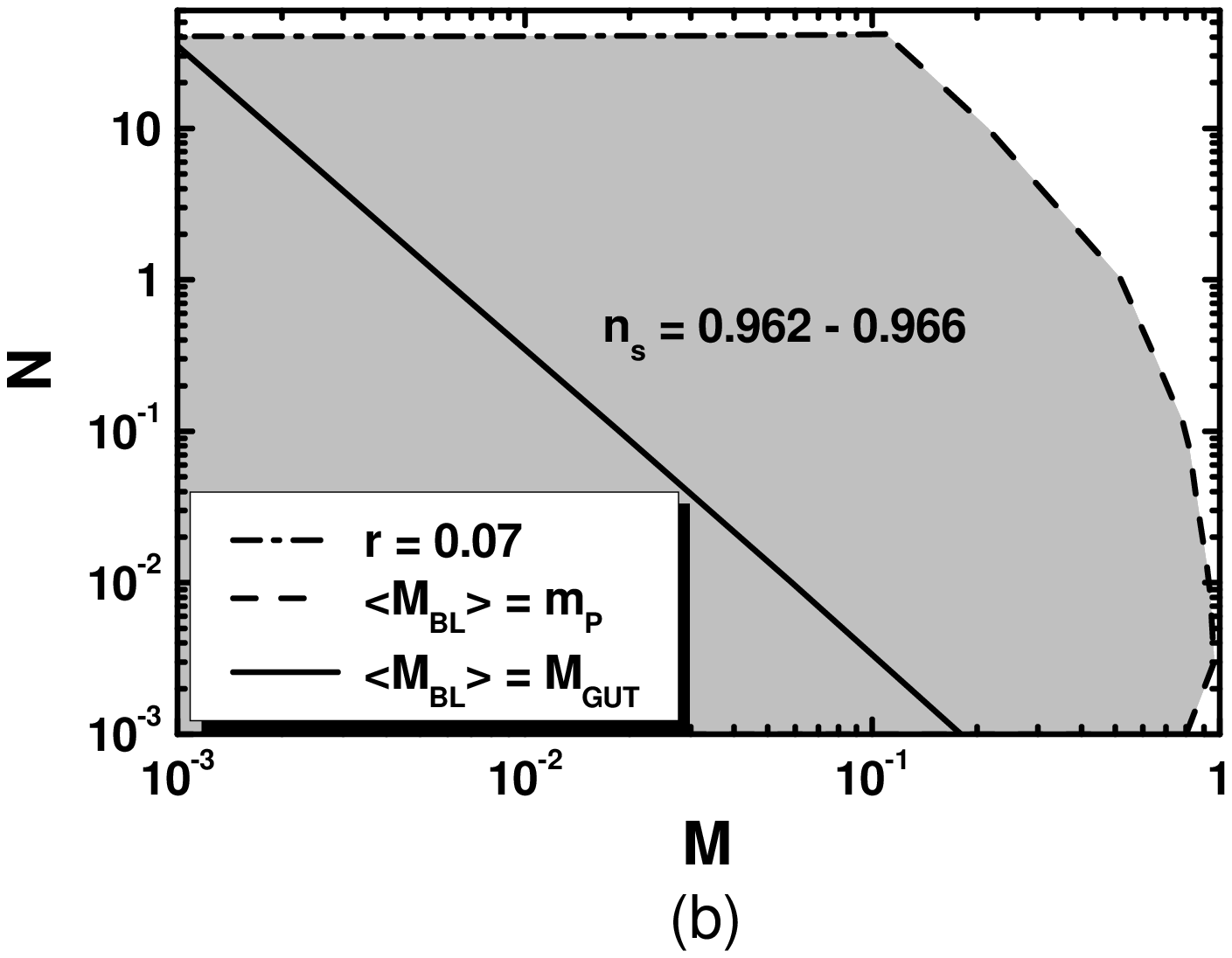,height=3.6in,angle=-90} \hfill
\end{minipage}
\hfill \vchcaption[]{\sl\small Allowed (shaded) regions in the
{\sffamily\ftn (a)} $M-(-\rs)$ plane for MI and {\sffamily\ftn
(b)} $M-N$ plane for MII. The conventions adopted for the various
lines are also shown.}\label{fig3}
\end{figure}

On the other hand, the allowed (shaded) region for \mb\ is
delineated in the $M-N$ plane. Here we extend our scanning to
non-integer $N$ too. We impose an artificial lower bound on $N$
stemming from naturalness. In fact, the tuning respecting $\Dex$
is worsened as $N$ decreases and so enforcing $\Dex\geq10^{-6}$
applies $N\geq10^{-3}$. The upper boundary curve of this region
comes from the bound on $r$ in \Eref{nswmap} which is depicted by
a dot-dashed line and provides an (almost constant) upper bound on
$N$. The rightmost border of the shaded region, represented by a
dashed line, originates from the saturation of the expectation
that $\vev{\mbl}\leq\mP$. Along the solid line we obtain the
theoretically better motivated value of $M$ corresponding to
$\vev{\mbl}=\mgut$. Taking the logarithm of both sides of
\Eref{Mg} with constant $g$, we can easily confirm the linear
dependence of $\ln N$ on $\ln M$ with negative coefficient, in
accordance with the depiction of the solid line in
\sFref{fig3}{b}. In the whole allowed region, we obtain
\beqs\bea\label{resm2}  10^{-3}\lesssim M\lesssim0.521,\>\>\>
10^{-3}\lesssim {N}\lesssim40,\>\>\>0.962\lesssim \ns\lesssim0.966,\>\>\>2.9\cdot10^{-6}\lesssim r\lesssim0.07;\>\>\>\\
5.3\cdot 10^{-3}\lesssim \ld/10^{-4}\lesssim3.7,\>\>\>
10^{-6}\lesssim {\Dex}\lesssim0.136\>\>\>
\mbox{and}\>\>\>0.3\lesssim
{\msn}/10^{11}~\GeV\lesssim721\,.\eea\eeqs Meanwhile, $\as$ is
confined in the range $-(6.1-7.4)\cdot10^{-4}$. We remark a
stronger variation of $\ld$, $\Dex$ and $\msn$ here from that
shown below \Eref{resm1} due to the variation of $N$ which
drastically affects the parameters above -- see \eqs{Nhi2}{lan2}.
Finally, since $\Dex$ changes with $N$ its upper bound remains
intact from the $M$ alteration and coincides with that in
\Eref{res2}.


\section{Conclusions}\label{con}

We established, in the context of SUGRA, a new type of HI i.e.,
inflation employing as inflaton $\sg$ the radial components of a
conjugate pair of Higgs fields. To be specific, in the current
paper these Higgs fields implement the breaking of a gauge
$U(1)_{B-L}$ symmetry at a scale $M$ which may assume a value
compatible with MSSM unification. This is realized making use of a
superpotential $W$ determined by an $R$ and the gauge symmetries
and containing the first allowed non-renormalizable term -- see
\Eref{Whi}. We judiciously selected two pairs of \Ka s, the
``untilded" and the ``tilded" ones -- see \eqs{Ks}{tKs} --, which
respect the symmetries above, parameterize hyperbolic spaces --
see \Eref{mnfs} -- and give rise to two models of HI, \ma\ and \mb
-- see \Eref{ms}. Both models share the same function which
relates the initial, $\sg$, with the canonically normalized
inflaton, $\se$ and includes a pole of order two -- see \Eref{Je}.

MI is relied on $W$ and the ``untilded" $K$'s which allow the pole
to appear also in the SUGRA potential $\Vhi$. The denominator of
$\Vhi$ though almost cancels out by constraining the scalar
curvatures of the corresponding \Km s in \Eref{Rs} to the values
${\cal R}_{(11)^2}=-4$ or ${\cal R}_{21}=-3$ and the coefficients
of the $W$ terms via the parameter $\rs$ in \Eref{rs}. In sharp
contrast, the pole remains invisible in $\Vhi$ of MII which is
based on the more structured, ``tilded" $K$'s and employ only
renormalizable terms in $W$. In this case, the scalar curvatures
of the corresponding \Km s assume the values ${\cal
R}_{(11)^2}=-4/N$ or ${\cal R}_{21}=-3/N$ with variable $N$.

Both models excellently match the observations by restricting the
free parameters to reasonably ample regions of values. In
particular, within \ma\ any observationally acceptable $\ns$ is
attainable by tuning $\rs$ to values of the order $10^{-5}$,
whereas $r$ is kept at the level of $10^{-3}$ -- see \Eref{res1}.
On the other hand, \mb\ avoids any tuning, larger $r$'s are
achievable as $N$ approaches its maximal allowed value $40$, while
$\ns$ lies close to its central value -- see \Eref{res2}. The
inflaton mass is collectively confined to the range
$(1.7-2.1)\cdot10^{11}~\GeV$. Varying $M$ beyond its SUSY value,
we obtain larger $\msn$'s and an mitigation of the tuning above
within \ma. In all cases, HI is realized with \sub\ values of
$\sg$, stabilizing thereby our predictions from possible higher
order terms in $W$, and can remain immune from the one-loop
radiative corrections.


Finally, we would like to point out that, although we have
restricted our discussion on the $\Ggut=G_{\rm SM}\times
U(1)_{B-L}$ gauge group, HI analyzed in this paper has in fact a
general character and can be applied in many extensions of the
MSSM. It can be realized within other GUTs, provided that $\bar
\Phi$ and $\Phi$ consist a conjugate pair of Higgs superfields. If
we adopt another GUT gauge group, the inflationary predictions are
expected to be quite similar to the ones discussed here with
possibly different analysis of the stability of the inflationary
trajectory, since different Higgs superfield representations may
be involved in implementing the $\Ggut$ breaking to $G_{\rm SM}$.
Removing the scale $M$ from $W$ in \Eref{Whi} and abandoning the
idea of grand unification, our HI can be realized even by the
electroweak Higgs boson within next-to-MSSM -- cf. \cref{sm1}. In
a such case the electroweak symmetry breaking can be processed
taking into account the soft-SUSY breaking terms. Since our main
aim here is just the establishment of the pole-induced HI, we
opted to utilize the simplest possible GUT embedding. Within
non-SUSY framework, MII can be also implemented within non-linear
sigma models which share similar \Kaa geometries. In such a case
the potential, put by hand, is given by \Eref{Vhi} swapping
$\ld^2$ with $\ld$. As a consequence, the tuning on $\ld$ is
aggravated since we expect $\ld\sim10^{-8}$ -- see $\ld$ values in
the Table of \Fref{fig1}. Moreover, in the absence of SUSY, the
gauge hierarchy problem -- somehow addressed in SUSY frameworks --
requires a special treatment.

\paragraph*{\small \bf\scshape Acknowledgments} {\small I would like to thank G. Lazarides
for useful discussions. This research work was supported by the
Hellenic Foundation for Research and Innovation (H.F.R.I.) under
the ``First Call for H.F.R.I. Research Projects to support Faculty
members and Researchers and the procurement of high-cost research
equipment grant'' (Project Number: 2251).}

\newpage
\setcounter{section}{0} \setcounter{subsubsection}{0}
\setcounter{equation}{0}
\renewcommand{\theequation}{A.\arabic{equation}}
\renewcommand{\thesubsubsection}{A.\arabic{subsubsection}}
\renewcommand{\thesubsection}{A.\arabic{subsection}}
\renewcommand{\thesection}{\alphabet{section}}

\appendix
\section*{Appendix A: Mathematical Supplement}\label{math}


We review here some mathematical properties regarding the
geometrical structure of $\phcb-\phc$ moduli space -- the geometry
of the $S$ dependent part of the adopted $K$'s is analyzed in
\cref{su11}. We focus our attention on $\kaa, \tkaa$ in \Sref{kaa}
and $\kba, \tkba$ in \Sref{kba}.

Let us, initially, recall that Riemannian metric, defined by the
line element $ds_K^2$, and the scalar curvature, ${\cal R}_{K}$,
associated with each of $K$'s above are calculated employing the
standard formulae \cite{gref}
\beq \label{ds} ds_K^2=K_{I\bar J}dZ^I dZ^{*\bar
J}\>\>\>\mbox{and}\>\>\>{\cal R}_{K}=-K^{I\bar
J}\partial_I\partial_{\bar J}\ln\lf\det M_{\phcb\phc}\rg\eeq
where $M_{\phcb\phc}=\lf K_{I\bar J}\rg$ expresses the kinetic
mixing in the inflationary sector. It is defined consistently with
\Eref{Kab} but it is here  computed without taking into account
the constraint of \Eref{inftr}.

\rhead[\fancyplain{}{ \bf \thepage}]{\fancyplain{}{\sl
Pole-Induced HI with Hyperbolic K\"ahler Geometries}}
\lhead[\fancyplain{}{\sl \hspace*{-0.1cm} Appendix A: Mathematical
Supplement}]{\fancyplain{}{\bf \thepage}} \cfoot{}

\subsubsection{The \Kaa Potentials $\kaa$ and $\tkaa$}\label{kaa}

Applying \Eref{ds} for $K=\kaa$ and $\tkaa$ we find
\beq\label{dsa}
ds_{(11)^2}^2=\frac{2N|d\phc|^2}{\lf1-2|\phc|^2\rg^{2}}+\frac{2N|d\phcb|^2}
{\lf1-2|\phcb|^2\rg^{2}}\>\>\>\mbox{and}\>\>\>{\cal
R}_{(11)^2}=-\frac{4}{N},\eeq
where $ds_{(11)^2}^2$ has a diagonal structure and is well defined
imposing the restrictions $|\phc|<1/\sqrt{2}$ and
$|\phcb|<1/\sqrt{2}$. Each of the two terms of $ds_{(11)^2}^2$
represents the extensively analyzed \cite{sky} Poincar\'e disc
which is the non-compact two-dimensional manifold $SU(1,1)/U(1)$
with scalar curvature equal to one half of ${\cal R}_{(11)^2}$
\cite{su11}. It is well-known that $SU(1,1)/U(1)$ is also
parameterized in the half-plane coordinates (usually notated by
$T$ and $T^*$) which are related to the disc coordinates -- e.g.,
$\phc$ and $\phc^*$ -- through a Cayley transformation
\cite{tkref, 7disk, alinde}. This parameterization, though,
violates the gauge symmetry and so it is unappropriate for our
purposes.

An element of the coset space $SU(1,1)/U(1)$ may be represented as
\cite{su11,class}
\beq
\label{cU}\cU=\mtta{\alpha}{b}{b^*}{\alpha}\>\>\>\mbox{with}\>\>\>\alpha\in\mathbb{R}_+
,\>\>b\in\mathbb{C}\>\>\>\mbox{and}\>\>\>\alpha^2-|b|=1.\eeq
Indeed, taking into account the definition of the coset space, we
can convince ourselves that the matrix \beq U=\cU
\ch=\mtta{a}{c}{c^*}{a^*}\>\>\>\mbox{with}\>\>\>\ch=\diag\lf
e^{i\vth}, e^{-i\vth}\rg,\>\>\> a=\alpha e^{i\vth}
\>\>\>\mbox{and}\>\>\> c=be^{i\vth},\eeq is an element of
$SU(1,1)$ satisfying the definitive relations
\beq \label{u11} U^\dag \eta_{11}
U=\eta_{11}\>\>\>\mbox{and}\>\>\>\det
U=1\>\>\>\mbox{with}\>\>\>\eta_{11}=\diag\lf1,-1\rg, \eeq
since $|a|^2-|c|^2=1$. We can consider two elements $\cU_1\in
SU(1,1)/U(1)_{\phc}$ and $\cU_2\in SU(1,1)/U(1)_{\phcb}$ acting on
$\phc$ and $\phcb$ respectively according to the linear fractional
transformations
\beq \sqrt{2}\phc\to \frac{\alpha_1
\sqrt{2}\phc+b_1}{b_1^*\sqrt{2} \phc +\alpha_1}
\>\>\>\mbox{and}\>\>\> \sqrt{2} \phcb\to \frac{\alpha_2\sqrt{2}
\phcb+b_2}{b_2^*\sqrt{2} \phcb+\alpha_2}\,, \label{tr11}\eeq where
$\alpha_i^2-|b_i|^2=1$ with $i=1,2$ and the elements $\alpha_i$
and $b_i$ are contained in $\cU_i$. The transformations above do
not violate the $\Gbl$ symmetry, provided that the elements of
$\cU_i$ are charged under $\Gbl$ as follows
\beq \label{bla} (\bl)(\alpha_1, b_1,\alpha_2, b_2 )=(0,1,0,-1).
\eeq
It is straightforward to show that $ds_{(11)^2}^2$ in \Eref{dsa}
remains invariant under the transformations in \Eref{tr11} and so
we conclude that $\kaa$ and $\tkaa$ parameterize $\mnfa$.

It is well-known \cite{gref} that a SUGRA model is fully described
by the function $G$, see \Eref{G}, which remains invariant under a
\Kaa transformation
\beq \label{Kst} K\to
K+\Lambda+\Lambda^*\>\>\>\mbox{and}\>\>\>W\to We^{-\Lambda}\,.\eeq
Therefore, models described by $K$'s and $W$'s related by such a
transformation are equivalent. We can prove that $\kaa$ in
\Eref{Ks} remains invariant, in the sense of \Eref{Kst}, i.e., it
is transformed as in \Eref{Kst} with
\beq \label{La} \Lambda=N\ln
\lf(b_1^*\sqrt{2}\phc+\alpha_1)(b_2^*\sqrt{2}\phcb+\alpha_2)\rg\,.
\eeq
Similar transformation is valid for $\tkaa$ in \Eref{tKs} since it
is related to $\kaa$ via another \Kaa transformation.

\subsubsection{The \Kaa Potentials $\kba$ and $\tkba$}\label{kba}

For $K=\kba$ and $\tkba$, \Eref{ds} yields
\beq\label{dsb}
ds_{21}^2=2N\lf\frac{|d\phc|^2+|d\phcb|^2}{1-|\phc|^2-|\phc|^2}+
\frac{|\phc^*d\phc+\phcb^*d\phcb|^2}
{\lf1-|\phc|^2-|\phcb|^2\rg^{2}}\rg\>\>\>\mbox{and}\>\>\>{\cal
R}_{21}=-\frac{3}{N},\eeq
where $ds_{21}^2$ is known in the mathematical literature -- see
e.g. \cref{math1} -- as the Bergmann metric defined in the open
ball $|\phc|^2+|\phcb|^2<1$ of $\mathbb{C}^2$. This is the
non-compact, four-dimensional manifold $\mnfb$ in \Eref{mnfs}. The
value of ${\cal R}_{21}$ above is in accord with the generic
result found in \cref{eno19}, if we take into account the
different sign adopted in the definition of ${\cal R}_K$ in
\Eref{ds} -- recall that the transition between the two notations
is realized setting $2N=3\alpha$. As for the case of $\mnfa$,
alternative parameterizations of $\mnfb$ \cite{eno7,class, eno7n}
are not allowed in our case, since these are not consistent with
the gauge symmetry.

We can find a representative element of $\mnfb$ following the
method presented in \cref{math}. Namely, an element of $SU(2,1)$
satisfying the definitive relations
\beq\label{su21} U^\dag \eta_{21}
U=\eta_{21}\>\>\>\mbox{and}\>\>\>\det
U=1\>\>\>\mbox{with}\>\>\>\eta_{21}=\diag\lf1,1,-1\rg, \eeq
may be written as \beq\label{cUb} U=\cU \ch \>\>\>\mbox{with}
\>\>\>\cU=\mttb{1/\na}{0}{a}{\na b a^*}{\na\gamma}{b}{\na \gamma
a^*}{\na b^*}{\gamma} \>\>\>
\>\>\>\mbox{and}\>\>\>\ch=e^{i\vth}\mttb{d
}{f}{0}{-f^*}{d^*}{0}{0}{0}{e^{-3i\vth}},\eeq where
$\na=1/\sqrt{1+|a|^2}$ and the free parameters $a,b,\gamma,d$ and
$f$ are constrained as follows
\beq a,b,d,f\in \mathbb{C}, \gamma\in
\mathbb{R_+}\>\>\>\mbox{with}\>\>\>
|a|^2+|b|^2-\gamma^2=-1\>\>\>\mbox{and}\>\>\>|d|^2+|f|^2=1.\eeq
The former constraint ensures the hyperbolic structure of $\mnfb$
whereas the latter indicates the compact form of $SU(2)$ -- cf.
\cref{su11}. Obviously, $\ch\in SU(2)\times U(1)$ -- which is
subgroup of $SU(2,1)$ -- and depends on four (real) parameters.
Consequently, $\cU$ is a representative of $\mnfb$ depending on
four parameters. It is trivial to verify that the constraints in
\Eref{su21} are fulfilled for $U=\cU$ and $U=\ch$ and so for
$U=\cU \ch$ too. As a cross check, recall that the pseudo-unitary
group $SU(2,1)$ depends on eight (4+4) free parameters -- cf.
\cref{class}.

The operation of $\cU\in \mnfb$ on $\phc$ and $\phcb$ can be
represented via the isometric transformations -- cf. \Eref{tr11}
\beq \phc\to \frac{(1/\na)\phc+\na b^*a\phcb+\na a\gamma}{a^* \phc
+b^* \phcb+\gamma} \>\>\>\mbox{and}\>\>\> \phcb\to \frac{\na
\gamma \phcb+\na b}{a^* \phc +b^* \phcb+\gamma}\,,
\label{tr21}\eeq where the $\phc$ and $\phcb$ independent
parameters originate from the lines of $\cU^\dag$. To keep
consistency with the $\Gbl$ symmetry we assign to the free
parameters of $\cU$ the $\bl$ charges
\beq \label{blb} (\bl)(a, b,\gamma)=(1,-1,0). \eeq
It is straightforward to show that $ds_{21}^2$ in \Eref{dsb}
remains invariant under \Eref{tr21} and so, we conclude that
$\kba$ and $\tkba$ parameterize $\mnfb$. Moreover, $\kba$ in
\Eref{Ks} and $\tkba$ in \Eref{tKs} remain invariant under
\Eref{tr21}, up to a \Kaa transformation. E.g., $\kba$ is
transformed as $K$ in \Eref{Kst} with
\beq\label{Lb} \Lambda=2N\ln(a^* \phc +b^* \phcb+\gamma).\eeq
To our knowledge, the representation of $\cU$ in \Eref{cUb} and
the isometries in \Eref{tr21} are employed for the first time in
demonstrating the invariance of $ds_{21}^2$ and $\kba$ under
$SU(2,1)/(SU(2)\times U(1))$.

\def\ijmp#1#2#3{{\sl Int. Jour. Mod. Phys.}
{\bf #1},~#3~(#2)}
\def\plb#1#2#3{{\sl Phys. Lett. B }{\bf #1}, #3 (#2)}
\def\prl#1#2#3{{\sl Phys. Rev. Lett.}
{\bf #1},~#3~(#2)}
\def\rmp#1#2#3{{Rev. Mod. Phys.}
{\bf #1},~#3~(#2)}
\def\prep#1#2#3{{\sl Phys. Rep. }{\bf #1}, #3 (#2)}
\def\prd#1#2#3{{\sl Phys. Rev. D }{\bf #1}, #3 (#2)}
\def\prdn#1#2#3#4{{\sl Phys. Rev. D }{\bf #1}, no. #4, #3 (#2)}
\def\prln#1#2#3#4{{\sl Phys. Rev. Lett. }{\bf #1}, no. #4, #3 (#2)}
\def\npb#1#2#3{{\sl Nucl. Phys. }{\bf B#1}, #3 (#2)}
\def\npps#1#2#3{{Nucl. Phys. B (Proc. Sup.)}
{\bf #1},~#3~(#2)}
\def\mpl#1#2#3{{Mod. Phys. Lett.}
{\bf #1},~#3~(#2)}
\def\jetp#1#2#3{{JETP Lett. }{\bf #1}, #3 (#2)}
\def\app#1#2#3{{Acta Phys. Polon.}
{\bf #1},~#3~(#2)}
\def\ptp#1#2#3{{Prog. Theor. Phys.}
{\bf #1},~#3~(#2)}
\def\n#1#2#3{{Nature }{\bf #1},~#3~(#2)}
\def\apj#1#2#3{{Astrophys. J.}
{\bf #1},~#3~(#2)}
\def\mnras#1#2#3{{MNRAS }{\bf #1},~#3~(#2)}
\def\grg#1#2#3{{Gen. Rel. Grav.}
{\bf #1},~#3~(#2)}
\def\s#1#2#3{{Science }{\bf #1},~#3~(#2)}
\def\ibid#1#2#3{{\it ibid. }{\bf #1},~#3~(#2)}
\def\cpc#1#2#3{{Comput. Phys. Commun.}
{\bf #1},~#3~(#2)}
\def\astp#1#2#3{{Astropart. Phys.}
{\bf #1},~#3~(#2)}
\def\epjc#1#2#3{{Eur. Phys. J. C}
{\bf #1},~#3~(#2)}
\def\jhep#1#2#3{{\sl J. High Energy Phys.}
{\bf #1}, #3 (#2)}
\newcommand\jcapn[4]{{\sl J.\ Cosmol.\ Astropart.\ Phys.\ }{\bf #1}, no. #4, #3 (#2)}

\end{document}